\documentclass[aps,preprint,showpacs]{revtex4}

\usepackage[xdvi]{graphicx}
\usepackage{latexsym}
\usepackage{amsbsy}
\usepackage{amsmath}
\usepackage{amssymb}

\begin{document}

\date{\today}

\title{Derivation of a time dependent Schr\"odinger equation as
  quantum mechanical Landau-Lifshitz-Bloch equation}  

\author{R.\ Wieser}
\affiliation{1. International Center for Quantum Materials, Peking
  University, Beijing 100871, China \\  
2. Collaborative Innovation Center of Quantum Matter, Beijing 100871, China}

\begin{abstract}
 The derivation of the time dependent Schr\"odinger equation with
 transversal and longitudinal relaxation, as the
 quantum mechanical analog of the classical Landau-Lifshitz-Bloch 
 equation, has been described. Starting from the classical
 Landau-Lifshitz-Bloch equation the transition to quantum mechanics
 has been performed and the corresponding von-Neumann equation
 deduced. In a second step the time Schr\"odinger equation has been
 derived. Analytical proofs and computer simulations show the
 correctness and applicability of the derived Schr\"odinger equation.    
\end{abstract}

\pacs{75.78.-n, 75.10.Jm, 75.10.Hk}
\maketitle

\section{INTRODUCTION} \label{s:intro}
The Landau-Lifshitz equation \cite{landauPZS32} respectively
Landau-Lifshitz-Gilbert (LLG) equation \cite{gilbertIEEE04} are the
most prominent equations describing spin dynamics. These
equations are intensively used to describe any kind of magnetization
dynamics in ferromagnetic, antiferromagnetic or ferrimagnetic
materials with diameters from a few {\AA}ngstr{\"o}m (atomistic
description) to a micrometer length scale (micromagnetism). The
Landau-Lifshitz respectively LLG equation describes the motion of a
magnetic moment under the influence of an effective field
$\mathbf{H}_{\mathrm{eff}}$ which causes a precessional motion and an
additional friction (transversal relaxation) which leads to a parallel
alignment of the magnetic moment and the effective field. During
the relaxation the length of the magnetic moment is
conserved. However, there are situations, e.g. during ultrafast reversal
processes \cite{atxitiaPRB11} or the dynamics near the critical temperature
$T_C$ \cite{chubykaloPRB06}, where the description using the Landau-Lifshitz or
LLG equation fails because the magnetization is not necessarily
constant. This has been already pointed out by H. B. Callen in 1957
\cite{callenJPCS58}: the general equation of motion of a ferromagnetic material
has to be obtained by expanding the change of the magnetization
$\mathbf{M}$ in the three orthogonal vectors $\mathbf{M}$, 
$(\mathbf{M} \times \mathbf{H}_{\mathrm{eff}})$, and $\mathbf{M}
\times (\mathbf{M} \times \mathbf{H}_{\mathrm{eff}})$.
\begin{equation} \label{LLB}
\frac{\mathrm{d}\mathbf{M}}{\mathrm{d}t} = 
-\gamma \mathbf{M} \times
\mathbf{H}_{\mathrm{eff}}  - \gamma \alpha_{\mathrm{tr}} \mathbf{M}
\times \left(\mathbf{M} 
 \times \mathbf{H}_{\mathrm{eff}}\right) -  \gamma \alpha_{\mathrm{l}}
 \left(\mathbf{M} \cdot \mathbf{H}_{\mathrm{eff}} \right) \mathbf{M}
 \;.    
\end{equation}     
While $\gamma = g \mu_B/\hbar$ is the gyromagnetic ratio,
$\alpha_{\mathrm{tr}}$ and $\alpha_{\mathrm{l}}$ are scalar functions
of $\mathbf{M}$, and $\mathbf{H}_{\mathrm{eff}}$. This equation is
called Landau-Lifshitz-Bloch equation and without the last term
($\alpha_{\mathrm{l}} = 0$) equal to the Landau-Lifshitz equation. In
many cases this equation can be found for the corresponding magnetic
moment $\mathbf{m} = \mathbf{M}V$, where 
$V$ is the volume of the sample, or the normalized magnetic moment
$\mathbf{S} = \mathbf{m}/|\mathbf{m}|$. Depending on the
characteristics of the sample $\alpha_{\mathrm{tr}}$,
$\alpha_{\mathrm{l}}$ and $\mathbf{H}_{\mathrm{eff}}$ will be
different. In general $\mathbf{H}_{\mathrm{eff}}$ is given by the
negative gradient of the Hamiltonian $\cal H$ with respect to the
magnetization or magnetic moment e.g. $\mathbf{H}_{\mathrm{eff}} =
-\nabla_{\mathbf{M}} {\cal H}$, eventually modified by an additional stochastic
noise term $\bold{\xi}$ describe the influence of temperature
\cite{nowakARCP01} and further modification to take into account that
the magnetization and most of the  material parameters itself like the
anisotropy are temperature dependent \cite{nowakPRB05}.  

As for the effective field we can find for $\alpha_{\mathrm{tr}}$ and
$\alpha_{\mathrm{l}}$ different descriptions. The first proposal has
been given by H. B. Callen \cite{callenJPCS58}. The assumption there
is that the dissipative process is dominated by spin wave transitions from
$(\mathbf{k} = 0)$ to $(\mathbf{k} \neq 
0)$ where $\mathbf{k}$ is the wave vector of the spin wave. Callen
deduces for $\alpha_{\mathrm{tr}}$ and $\alpha_{\mathrm{l}}$ (Eq.~(36)
in \cite{callenJPCS58}):
\begin{eqnarray} 
\alpha_{\mathrm{tr}} = \frac{1}{\gamma |\mathbf{H}_{\mathrm{eff}}|}
\left[\frac{1}{n_0}\frac{\mathrm{d}n_0}{\mathrm{d}t} +
  \frac{2\gamma\hbar}{|\mathbf{M}|} \frac{\mathrm{d}n'}{\mathrm{d}t}
  \right] \,,\;  
\alpha_{\mathrm{l}} = \frac{\hbar}{|\mathbf{M}|}\frac{\mathrm{d}n'}{\mathrm{d}t}
\;, \nonumber
\end{eqnarray}
with $n' = \sum_{\mathbf{k}\neq 0} n_{\mathbf{k}}$ and $n_k =
a_{\mathbf{k}}^\dagger a_{\mathbf{k}}$, where $a_{\mathbf{k}}^\dagger$
and $a_{\mathbf{k}}$ are the Bose creation and annihilation
operators. 

R. S. Gekht et al. propose the following functions (see
Eq.~(3.4) in \cite{gekhtJETP76}):
\begin{eqnarray} 
\alpha_{\mathrm{tr}} = \alpha \,,\;\;\mathrm{and} \;\;
\alpha_{\mathrm{l}} = \frac{2 \alpha \gamma k_B T}{m}\;, \nonumber
\end{eqnarray}
with $\alpha$ the Gilbert damping constant, $k_B$ the
Boltzmann constant, $T$ the temperature and $m = |\mathbf{m}|$. The
assumption behind this proposal is the temperature dependence of the
magnetization. 

In 1990, the same idea following, D. A. Garanin 
et al. \cite{garaninTMP90,garaninPA91} and T.~Plefka
\cite{plefkaPA90,plefkaZPhysB93} proposed independently the following
functions for $\alpha_{\mathrm{tr}}$ and 
$\alpha_{\mathrm{l}}$ which can be found with just slightly
changes in nearly all recent publications
\cite{evansPRB12,chubykaloPRB06,haneyPRB09,xuPE12,xuJAP13} 
dealing with the Landau-Lifshitz-Bloch equation. Here in the writing
of L.~Xu and S.~Zhang \cite{xuPE12,xuJAP13}:
\begin{eqnarray} 
\alpha_{\mathrm{tr}} = \frac{m_{\mathrm{eq}}}{\gamma \tau_s m
  |\mathbf{H}_{\mathrm{eff}}|} \,,\; 
\alpha_{\mathrm{l}} = \frac{1}{\gamma
  \tau_s}\left[\frac{m}{\mathbf{m}\cdot\mathbf{H}_{\mathrm{eff}}} -
  \frac{m_{\mathrm{eq}}}{m|\mathbf{H}_{\mathrm{eff}}|}\right]\,. \nonumber
\end{eqnarray}
$m_{\mathrm{eq}} = |\mathbf{m}_{\mathrm{eq}}|$ is the
equilibrium magnetization and $\tau_s$ the spin relaxation time,
similar to $T_1$ and $T_2$ in the case of the Bloch equation
\cite{blochPR46}. In the most cases the temperature dependence of
$m_{\mathrm{eq}}$ is for simplicity reasons taken into account via a
mean field theory. 

In all these cases, even if the derivation starts
with a quantum mechanical description, the authors end up with the
(semi-) classical LLB equation [Eq.~(\ref{LLB})] where $\mathbf{M}$ is
either the magnetization $\mathbf{M}$, the magnetic moment
$\mathbf{m}$ or $\mathbf{S}$ or at least the spin expectation value
$\langle \hat{\mathbf{S}} \rangle$ of the spin operator
$\hat{\mathbf{S}}$. There are two reasons: The first reason can be seen in the 
Ehrenfest theorem \cite{ehrenfestZP27}, which says that the quantum
mechanical expectation values behave classical. However, in the mean
time it is known that the Ehrenfest theorem fails if the potential
$V(x)$ is not linear: $\langle x^a \rangle \neq \langle x \rangle^a$ if
$a \geq 2$. In the case of the Heisenberg model this means a classical
behavior of the spin expectation values can be expected only if the
terms of the Hamiltonian are linear in $\hat{\mathbf{S}}_n$, where $n$
is the spin index \cite{wieserPRB11}. This excludes especially crystalline
anisotropies which are proportional to $\hat{\mathbf{S}}_n^2$. 

The second reason is the fact that with the classical description larger system
sizes as with a quantum mechanical description can be addressed. This
can be explained with the fact that in the classical description the
spins are local: every spin can be addressed separately and is affected by a
local effective field. This makes it possible to simulate up to $10^6$
spins \cite{nowakARCP01}. In the quantum mechanical description we are
dealing with wave functions describing all spins at the same time. The
corresponding matrices are huge and actual it impossible to address
more than 64 spins $S = 1/2$ in maximum using exact diagonalization
\cite{sandvikAIPConf10}. The larger system sizes can be seen as an
advantage. On the other hand with a classical description quantum 
effects get lost. The comparison between classical and quantum spin
dynamics shows that a similar dynamics can be found only in some special
cases: 
\begin{description}
\item [1.] in the classical limit ($S \rightarrow \infty$, $\hbar
  \rightarrow 0$, and $\hbar S \rightarrow 1$) 
\item [2.] only linear terms in $\hat{\mathbf{S}}_n$, where
  $n$ is the lattice site, in the Hamilton operator $\hat{\mathrm{H}}$
  \cite{wieserPRB11}  
\item [3.] in the case of no entanglement \cite{wieserEPJB15}, e.g. if
  the system is described by a product state \cite{susskindBOOK}  
\item [4.] $|\psi\rangle$ corresponds to a
  superposition of the basis states $|S, m_S \rangle = |S, \pm S \rangle$ only: 
$|\psi \rangle = \psi_{+S} |S, +S \rangle + \psi_{-S} |S, -S \rangle$. 
\end{description}

The last scenario (point 4.) is the case for:
\begin{description}
\item [(a)] ferromagnetic spin waves: in this case $|\psi\rangle$ is
  approximately given by $|\psi\rangle \approx |S, \pm S \rangle$
  \cite{noltingBOOKMagE,kittelBOOKengl}
\item [(b)] coherent states where $|\psi\rangle$ is given by $|\psi\rangle
= U(\theta,\phi) |S, \pm S \rangle$, $U(\theta,\phi)$ is a unitary
transformation describing a rotation with the rotation angles $\theta$
and $\phi$ \cite{schliemannJPCONDMAT97}   
\item [(c)] a single spin with $S = 1/2$: in this case the wave function
  is always given by: 
  $|\psi\rangle = \psi_\uparrow |\uparrow\,\rangle
  + \psi_\downarrow |\downarrow\,\rangle$ (Bloch sphere)
  \cite{gisinHelvPhysActa81}  
\item [(d)] a single spin with $S > 1/2$ if the only contribution to
  $\hat{\mathrm{H}}$ is a external field in direction of the
  quantization axis (in the most cases $\mathbf{B} =
  B_z\hat{\mathbf{z}}$). Perpendicular fields lead to quantum
  tunneling which can lead to states $|\psi\rangle = |S,m_S\rangle$,
  with $m_S \neq \pm S$
  \cite{garciapablosJAP98,wernsdorferSCIENCE99,rastelliPRB01,wieserEPJB15}. 
\end{description}
As said before, the mentioned examples in the introduction using the
Landau-Lifshitz-Bloch equation describing a classical or semiclassical
spin dynamics which means they exclude quantum effects like
entanglement. The spin dynamics with or without entanglement is
totally different. The spin expectation values $\langle
\hat{\mathbf{S}} \rangle$ follow the same trajectories as the classical
spin $\mathbf{S}$ only if there is no entanglement
\cite{wieserEPJB15}. This together with the possibility to find
quantum tunneling in anisotropic spin systems
\cite{garciapablosJAP98,wernsdorferSCIENCE99,rastelliPRB01} are the
main differences between the quantum mechanical description which
takes these effects into account and the classical or a semiclassical
description which do not take into account these effects. 
  
To take into account these quantum effects it is necessary to describe the
system fully quantum mechanical and to calculate the spin expectation
values at the end. The goal of this publication is to give a time
dependent Schr\"odinger equation which enables us to address all
quantum effects and at the same time to take into account transversal
and longitudinal relaxation similar to the (semi-) classical
description using the LLB equation [Eq.~(\ref{LLB})]. 

The outline of the publication is the following:
In Sec.~\ref{s:LLB} first the von Neumann equation will be
introduced and after that the corresponding time dependent Schr\"odinger
equation will be derived. The reason for this is the facts
that the von Neumann equation is easier to understand and closer to
the (semi-) classical description than the time dependent
Schr\"odinger equation. However, the time dependent
Schr\"odinger equation has a reduced numerical effort with respect to
the von Neumann equation. For a Hilbert space of dimension $N$ the
number of components of the corresponding wave 
function $|\psi\rangle$ is $N$ while the number of matrix components
of the density operator matrix $\hat{\rho} = |\psi\rangle \langle \psi
|$ is equal to $N^2$ \cite{molmerJOSAB93}. In Sec.~\ref{s:proof} the
derived time dependent Schr\"odinger equation will be proved
analytical. Here the Hamiltonian is chosen in such a way that the we
can expect a classical behavior of the spin expectation value $\langle
\hat{\mathbf{S}} \rangle$. This gives us a direct proof of the
correctness of our description. Sec.~\ref{s:num} demonstrates the
possibility to solve the derived time dependent Schr\"odinger equation
under some more complex condition and the stability of the numerical
calculation. The publication ends with a summary (Sec.~\ref{s:summary}).

\section{EQUATION OF MOTION} \label{s:LLB} 
In a recently published manuscript \cite{wieserEPJB15} it has been  
shown that the following von Neumann equation: 
\begin{equation} \label{LiouvilleOriginal}
\frac{\mathrm{d}\hat{\rho}}{\mathrm{d}t} =
\frac{i}{\hbar}[\hat{\rho},\hat{\mathrm{H}}] -
  \frac{\alpha_{\mathrm{tr}}}{\hbar}[\hat{\rho},[\hat{\rho},\hat{\mathrm{H}}]]
  \; 
\end{equation} 
and the corresponding self-consistent nonlinear Schr\"odinger equation:
\begin{equation} \label{TDSEOriginal}
i \hbar \frac{\mathrm{d}}{\mathrm{d}t} | \psi\rangle =
\hat{\mathrm{H}}|\psi \rangle - i \alpha_{\mathrm{tr}} \left(\hat{\mathrm{H}} -
  \langle\hat{\mathrm{H}}\rangle\right) |\psi \rangle \;,
\end{equation}
with $\langle \hat{\mathrm{H}} \rangle = \langle \psi
|\hat{\mathrm{H}}| \psi \rangle$, and $\hat{\mathrm{H}}$ an arbitrary
Hermitian Heisenberg Hamiltonian, can be used to describe the dynamics
of quantum spin. In both equations the first term on the right hand
side describes a precessional motion and the second term a transversal
relaxation.  

It has been further shown that in the case of a spin
$\hat{\mathbf{S}}_n$, where $n$ corresponds to the $n$th spin at 
lattice site $\mathbf{r}_n$, in an effective field $\mathbf{B}_n$ the
trajectories of the spin expectation value $\langle \hat{\mathbf{S}}_n
\rangle = \langle \psi |\hat{\mathbf{S}}_n|\psi \rangle$, where the
wave function $|\psi \rangle$ has been calculated with
Eq.~(\ref{TDSEOriginal}), are similar to  
the trajectories of the classical spin $\mathbf{S}_n$ with
dynamics described by the Landau-Lifshitz equation:
\begin{equation} \label{LLOri}
\frac{\mathrm{d}\langle \hat{\mathbf{S}}_n \rangle}{\mathrm{d}t} = 
\gamma \langle \hat{\mathbf{S}}_n \rangle \times
\mathbf{B}_n  - \gamma \alpha_{\mathrm{tr}} \langle \hat{\mathbf{S}}_n \rangle
\times \left(\langle \hat{\mathbf{S}}_n \rangle 
 \times \mathbf{B}_n\right) \;.    
\end{equation}   
The only difference between both descriptions is the reversed sense of
rotation of the precessional motion (first term: $\gamma \rightarrow
-\gamma$).  

However, there are two restrictions: Eq.~(\ref{LLOri}) holds only if
the Hamiltonian is linear in $\hat{\mathbf{S}}_n$ (point 2 in the list
before). This means as long as we can write $\hat{\mathrm{H}} =
-\sum_n \mathbf{B}_n \cdot \hat{\mathbf{S}}_n$. Thereby, the effective
field $\mathbf{B}_n$ itself can be a function of the surrounding spins
$\hat{\mathbf{S}}_m$, $m \neq n$, interacting with
$\hat{\mathbf{S}}_n$. Higher order contributions as uniaxial 
anisotropies or the biquadratic exchange (both quadratic in
$\hat{\mathbf{S}}_n$), will lead to additional terms of the order
$\hbar$, which disappear in the classical limit $\hbar
\rightarrow 0$ \cite{wieserPRB11}.  

The second restriction is related to the entanglement. While
Eq.~(\ref{LLOri}) looks similar to the classical Landau-Lifshitz
equation the trajectories of spin expectation values $\langle
\hat{\mathbf{S}}_n \rangle$ and the corresponding classical spin
$\mathbf{S}_n$ are not necessary equal. The trajectories differ if
entanglement plays a role. While in the (semi-) classical description
the length of the spin $|\mathbf{S}_n|$ only changes if there is an 
additional longitudinal relaxation as in the Landau-Lifshitz-Bloch
equation the value $|\langle \hat{\mathbf{S}}_n \rangle|$ can also
change if there is no longitudinal relaxation in the equation of
motion as in Eq.~(\ref{LLOri}). Indeed, $|\langle \hat{\mathbf{S}}_n
\rangle|$ is a direct indicator for entanglement: $|\langle \hat{\mathbf{S}}_n
\rangle| = \hbar S$ can be expected only if the system shows no
entanglement \cite{wieserEPJB15}. In all the other cases we have
$|\langle \hat{\mathbf{S}}_n \rangle| < \hbar S$. In a case of maximal
entanglement as in the case of singlet state
$|\mathrm{singlet} \rangle = (|\uparrow \downarrow \rangle -
|\downarrow \uparrow \rangle)/\sqrt{2}$ we have $|\langle \hat{\mathbf{S}}_n 
\rangle| = 0$ \cite{susskindBOOK}.   

To take the entanglement into account it is mandatory necessary that
we solve the time dependent Schr\"odinger Eq.~(\ref{TDSEOriginal}) or
the corresponding von Neumann Eq.~(\ref{LiouvilleOriginal}) and in a
second step calculate the spin expectation values $\langle
\hat{\mathbf{S}}_n \rangle$. Eq.~(\ref{LLOri}) does 
not take into account the change of $|\langle \hat{\mathbf{S}}_n
\rangle|$ and therefore entanglement. This means the description where
we solve Eq.~(\ref{LLOri}) to calculate the trajectories of $\langle
\hat{\mathbf{S}}_n \rangle$ has to be considered as semiclassical. The
resulting trajectories in this case will be the same as for the
classical Landau-Lifshitz equation but not necessarily the same as in
the case of the quantum mechanical description. The same is true for
the Landau-Lifshitz-Bloch equation and the corresponding Schr\"odinger
equation which we will derive in the following.     

The first step will be to derive the corresponding von Neumann
equation. As going from the Landau-Lifshitz equation
[Eq.~(\ref{LLOri})] to the Landau-Lifshitz-Bloch 
equation [Eq.~(\ref{LLB})] we have to add to the von Neumann equation
[Eq.~(\ref{LiouvilleOriginal})] an additional
longitudinal relaxation term which can be derived from the
longitudinal relaxation term of the classical Landau-Lifshitz-Bloch
equation: $\gamma
\alpha_{\mathrm{l}} \left(\mathbf{S} \cdot \mathbf{B} \right)
\mathbf{S}$ (we assume $\mathbf{H}_{\mathrm{eff}} = \mathbf{B}$)
in the following way: we replace the classical spin $\mathbf{S}$ by the 
expectation value $\langle \hat{\mathbf{S}} \rangle$ and correct the
dimension. $\mathbf{S}$ is assumed to be dimensionless, but
$\langle \hat{\mathbf{S}} \rangle$ has the dimension of $\hbar$. We
correct the dimension by an additional $1/\hbar$ to keep
$\alpha_{\mathrm{l}}$ dimensionless:  
\begin{equation} \label{LongJohnSilver}
-\gamma \alpha_{\mathrm{l}} \left(\mathbf{S} \cdot \mathbf{B} \right)
\mathbf{S} \rightarrow
-2\frac{\alpha_{\mathrm{l}}}{\hbar}\left(\frac{g\mu_B}{\hbar}\mathbf{B}
\cdot \langle  
\hat{\mathbf{S}}\rangle \right) \langle \hat{\mathbf{S}} \rangle = 
2\frac{\alpha_{\mathrm{l}}}{\hbar} \langle
\hat{\mathrm{H}} \rangle \langle \hat{\mathbf{S}} \rangle \;.\nonumber
\end{equation}
The same dimension problem also appears for $\alpha_{\mathrm{tr}}$. We
have to correct the dimension there too. The additional factor 2 is
needed to guarantee later a symmetric decoupling during the derivation of
the time-dependent Schr\"odinger equation. Furthermore, it can be
shown that this factor is needed to get the correct spin length (see
supplementary material \cite{Suppl}).

The next step is to write $\langle \hat{\mathrm{H}} \rangle \langle
\hat{\mathbf{S}} \rangle$ as:
\begin{equation} \label{HSImportExport}
 \langle
\hat{\mathrm{H}} \rangle \langle \hat{\mathbf{S}} \rangle =
\mathrm{Tr}(\hat{\rho}\hat{\mathrm{H}})\mathrm{Tr}(\hat{\rho}\hat{\mathbf{S}})
= \mathrm{Tr}(\hat{\rho}\hat{\mathrm{H}}\hat{\rho}\hat{\mathbf{S}}) \;
\end{equation} 
\cite{wieserEPJB15}. The relaxation term in terms of the density
operator $\hat{\rho}$ appears if we skip $\mathrm{Tr}$ and
$\hat{\mathbf{S}}$ in Eq.~(\ref{HSImportExport}): 
$\mathrm{Tr}(\hat{\rho}\hat{\mathrm{H}}\hat{\rho}\hat{\mathbf{S}})
\rightarrow \hat{\rho}\hat{\mathrm{H}}\hat{\rho}$. Adding the
resulting expression to Eq.~(\ref{LiouvilleOriginal}) leads to the von Neumann
equation containing transversal and longitudinal relaxation:
 \begin{equation} \label{Liouville}
\frac{\mathrm{d}\hat{\rho}}{\mathrm{d}t} =
\frac{i}{\hbar}[\hat{\rho},\hat{\mathrm{H}}] -
  \frac{\alpha_{\mathrm{tr}}}{\hbar}[\hat{\rho},[\hat{\rho},\hat{\mathrm{H}}]]
  + 2\frac{\alpha_{\mathrm{l}}}{\hbar}
  (\hat{\rho}\hat{\mathrm{H}})\hat{\rho} \;.  
\end{equation} 

To derive the corresponding Schr\"odinger
equation we use the definition of the density operator in the case of
a pure state:
\begin{equation} \label{Rho}
\hat{\rho} = |\psi \rangle\langle \psi| \;. 
\end{equation}
Inserting $\hat{\rho}$ in Eq.~(\ref{Liouville}) we find
after some algebra the following two differential
equations:   
\begin{eqnarray} \label{ZwischenEqWichtig1}
\frac{\mathrm{d}|\psi\rangle}{\mathrm{d}t}\langle \psi|   
&=&  \left[-\frac{i}{\hbar}\hat{\mathrm{H}}|\psi \rangle -
  \frac{\alpha_{\mathrm{tr}}}{\hbar}\left (\hat{\mathrm{H}} - \langle 
    \hat{\mathrm{H}} \rangle \right)|\psi \rangle
    \right]\langle \psi | \nonumber \\ 
&+& \frac{\alpha_{\mathrm{l}}}{\hbar} |\psi\rangle \langle 
    \hat{\mathrm{H}} \rangle \langle \psi |  \;
\end{eqnarray}
\begin{eqnarray} \label{ZwischenEqWichtig2}
|\psi\rangle \frac{\mathrm{d}\langle \psi|}{\mathrm{d}t}  
&=&  |\psi \rangle \left[ \langle \psi |
  \hat{\mathrm{H}}\frac{i}{\hbar} - 
   \langle \psi | \left (\hat{\mathrm{H}} - \langle 
    \hat{\mathrm{H}} \rangle \right)
    \frac{\alpha_{\mathrm{tr}}}{\hbar} \right] \nonumber \\ 
&+& \frac{\alpha_{\mathrm{l}}}{\hbar} |\psi\rangle \langle 
    \hat{\mathrm{H}} \rangle \langle \psi | \;.
\end{eqnarray}

After multiplying Eq.~(\ref{ZwischenEqWichtig1}) with $|\psi \rangle$ from the
right and dividing both sides by $\langle \psi| \psi \rangle$ we find
the Schr\"odinger equation:  
\begin{equation} \label{TDSELLB}
i\hbar\frac{\mathrm{d}}{\mathrm{d}t} | \psi\rangle =
(\hat{\mathrm{H}} - 
  i\alpha_{\mathrm{tr}}[\hat{\mathrm{H}} -
  \langle\hat{\mathrm{H}}\rangle]  +
  i\alpha_{\mathrm{l}}  
    \langle\hat{\mathrm{H}}\rangle) |\psi \rangle\;.
\end{equation}
The three terms on the right hand side are the same as in the case
of the classical Landau-Lifshitz-Bloch equation. The first term
describes an undamped precession. The second term provides a
transversal and the last term a longitudinal relaxation. In the case
of the Landau-Lifshitz-Bloch equation the transversal and longitudinal
relaxation act separate, which means independent. However, a careful
analysis shows that this is not the case for the recently proposed
Schr\"odinger equation. This equation has a problem if we assume a
spin oriented parallel to an external field $\mathbf{B} = 
B_z \hat{\mathbf{z}}$. In this case only the longitudinal relaxation should  
contribute and lead to a change the length of the spin. A transversal
relaxation should taking place. However, it can be shown that in this
simple scenario both relaxation terms influence the change of the spin
length. The influence of the transversal relaxation term can be seen
with $\hat{\mathrm{H}} = - g\mu_B B_z \hat{S}_z /\hbar = 
- b_z \hat{\sigma}_z$ and $|\psi(t) \rangle = \psi_\uparrow(t)|\uparrow
\,\rangle = \psi_\uparrow(t) (1,0)^T$ by:
\begin{equation}
-i\alpha_{\mathrm{tr}}[\hat{\mathrm{H}} -
\langle\hat{\mathrm{H}}\rangle] |\psi \rangle =
ib_z\alpha_{\mathrm{tr}}\psi_\uparrow(t) \left[1 -
  |\psi_\uparrow|^2\right]\left(\begin{array}{c}1\\0\end{array}\right)
  \nonumber \;. 
\end{equation}
This equation only becomes zero if $|\psi_\uparrow|^2 = \langle \psi |
\psi \rangle = 1$, which means in the case of a normalized wave
function. This is surely the case for $\alpha_{\mathrm{l}} = 
0$ (no longitudinal relaxation), where we deal with normalized wave functions
\cite{wieserEPJB15}. However, in the cases 
$\alpha_{\mathrm{l}} > 0$ we have a decrease or increase of the norm
of the wave function $n =  \langle \psi | \psi \rangle$ due to the 
longitudinal relaxation. In this case we have $0 \leq \langle \psi |
\psi \rangle \leq 1$ and therefore a contribution of the transversal
relaxation. This means we need a modification to fix this problem:  
\begin{equation} \label{TDSELLB2}
i\hbar\frac{\mathrm{d}}{\mathrm{d}t} | \psi\rangle =
(\hat{\mathrm{H}} - 
  i\alpha_{\mathrm{tr}}[\langle \psi | \psi \rangle
  \hat{\mathrm{H}} - 
  \langle\hat{\mathrm{H}}\rangle]  + 
  i\alpha_{\mathrm{l}}  
    \langle\hat{\mathrm{H}}\rangle)  |\psi \rangle\;.
\end{equation}
The modification is to add $\langle \psi | \psi \rangle$ into the
transversal relaxation term which leads to
$-i\alpha_{\mathrm{tr}}[\langle \psi | \psi \rangle \hat{\mathrm{H}} - 
\langle\hat{\mathrm{H}}\rangle ] |\psi \rangle = 0$ in any case. With
this modification the transversal relaxation does not contribute in
this constellation and we can expect the correct results.

\section{ANALYTICAL PROOF} \label{s:proof} 
It is easy to show that the conjugate transposed equation
corresponding to Eq.~(\ref{TDSELLB2}) is given by: 
\begin{equation} \label{ctTDSELLB2}
-i\hbar\frac{\mathrm{d}}{\mathrm{d}t} \langle \psi | =
 \langle \psi |( \hat{\mathrm{H}} + i 
  \alpha_{\mathrm{tr}} [
  \hat{\mathrm{H}} \langle \psi | \psi \rangle -
  \langle\hat{\mathrm{H}}\rangle ]  - i
  \alpha_{\mathrm{l}}  
    \langle\hat{\mathrm{H}}\rangle ) \;.
\end{equation}
Next, we are looking for the time development of a single spin
in an external field $\mathbf{B}$ described by:
\begin{equation}
\mathbf{m} = \frac{\langle \hat{\mathbf{S}} \rangle}{\hbar S} \;,
\end{equation}
with $\langle \hat{\mathbf{S}} \rangle = \langle \psi |
\hat{\mathbf{S}} | \psi \rangle$. 
Furthermore, we assume $S = 1/2$ which means $\hat{\mathbf{S}} =
\frac{\hbar}{2}\hat{\bold{\sigma}}$ and 
\begin{equation}
\mathbf{m} = \langle
  \psi |\hat{\bold{\sigma}} | \psi \rangle \;.
\end{equation} 
Therefore, the time derivative of $\mathbf{m}(t)$ is given by: 
\begin{equation} \label{mTime}
\frac{\mathrm{d}{\mathbf m}}{\mathrm{d} t} =
\langle \dot{\psi} 
|\hat{\mathbf{\sigma}} | \psi \rangle + \langle \psi
|\hat{\mathbf{\sigma}} | \dot{\psi} \rangle \;,
\end{equation}
where $| \dot{\psi} \rangle = \frac{\mathrm{d}}{\mathrm{d} t}|
\psi \rangle$ and $\langle \dot{\psi}| = \frac{\mathrm{d}}{\mathrm{d}
  t}\langle \psi|$ are represented by the time dependent
Schr\"odinger equation (\ref{TDSELLB2}) and the corresponding conjugate
transposed equation (\ref{ctTDSELLB2}).     

It is more convenient to look for one component of $\mathbf{m}$, e.g
$z$-component $m_z$. The time development of $m_z$ is given by:  
\begin{eqnarray} \label{MZTime}
\frac{\mathrm{d}m_z}{\mathrm{d} t} =  \langle \dot{\psi}
|\hat{\sigma}_z | \psi \rangle + \langle \psi
|\hat{\sigma}_z | \dot{\psi} \rangle \;.
\end{eqnarray}

Due to the assumption $S = 1/2$ the Hamilton operator of a single spin
in an external field is given by:  
\begin{equation} \label{HamHam}
\hat{\mathrm{H}} = -\frac{g\mu_B}{\hbar}\mathbf{B}\cdot\mathbf{\hat{S}} =
-\frac{g\mu_B}{2}\mathbf{B}\cdot\hat{\mathbf{\sigma}} \;, 
\end{equation}
and $\langle \hat{\mathrm{H}} \rangle$ 
as: 
\begin{equation} \label{HamHamExpect}
\langle \hat{\mathrm{H}} \rangle =
-\frac{g\mu_B}{2}\mathbf{B}\cdot \mathbf{m} \;.
\end{equation} 

Inserting the equations of motions Eq.~(\ref{TDSELLB2}) and
(\ref{ctTDSELLB2}), together with Eq.~(\ref{HamHam}) and
(\ref{HamHamExpect}) in Eq.~(\ref{MZTime}) we get after some algebra: 
\begin{eqnarray} \label{MZTimeII}
\frac{\mathrm{d}m_z}{\mathrm{d} t} =
&-&\frac{ig\mu_B}{2\hbar}\left(B_x\langle\psi|[\hat{\sigma}_x,\hat{\sigma}_z]|
\psi\rangle +
B_y\langle\psi|[\hat{\sigma}_y,\hat{\sigma}_z]|\psi\rangle \right)
\nonumber\\  
&+&\frac{g\mu_B\alpha_{\mathrm{tr}}}{2\hbar} \langle \psi | \psi
\rangle B_x\langle\psi|\{\hat{\sigma}_x, 
\hat{\sigma}_z\} |\psi\rangle  \nonumber \\
&+& \frac{g\mu_B\alpha_{\mathrm{tr}}}{2\hbar}
\langle \psi | \psi \rangle
B_y\langle\psi|\{\hat{\sigma}_y,\hat{\sigma}_z\}|\psi\rangle
\nonumber \\  
&+& \frac{g\mu_B\alpha_{\mathrm{tr}}}{\hbar} B_z [\langle\psi| \psi \rangle]^2
-\frac{g\mu_B(\alpha_{\mathrm{tr}} +
  \alpha_{\mathrm{l}})}{\hbar}(\mathbf{B}\cdot\mathbf{m})\, m_z  
\nonumber \\ 
\end{eqnarray}
Here, we have used the definition of $m_z = \langle \psi
|\hat{\sigma}_z|\psi \rangle$ and $\hat{\sigma}_z\hat{\sigma}_z =
\hat{\mathbf{1}}$, where $\hat{\mathbf{1}}$ is the identity matrix. The
same is true for the other Pauli matrices:
$\hat{\sigma}_x\hat{\sigma}_x = \hat{\sigma}_y\hat{\sigma}_y =
\hat{\mathbf{1}}$. In Eq.~(\ref{MZTimeII}) the
$[\hat{\sigma}_\alpha,\hat{\sigma}_\beta ] =
\hat{\sigma}_\alpha\hat{\sigma}_\beta -
\hat{\sigma}_\beta\hat{\sigma}_\alpha$ are commutators while the
$\{\hat{\sigma}_\alpha,\hat{\sigma}_\beta \} =
\hat{\sigma}_\alpha\hat{\sigma}_\beta +
\hat{\sigma}_\beta\hat{\sigma}_\alpha$ are
anticommutators. Independent of $S$ the commutators are given by
$[\hat{\sigma}_\alpha,\hat{\sigma}_\beta] = 
2i\epsilon_{\alpha,\beta,\gamma}\hat{\sigma}_\gamma$, where
$\epsilon_{\alpha,\beta,\gamma}$ is the Levi-Civita tensor. With changing
$S$, only the Pauli matrices change. This is not the case for the
anticommutators $\{\hat{S}_\alpha,\hat{S}_\beta \}$. They are changing
with $S$. In the case of $S = 1/2$ the anticommutators are 
given by: $\{\hat{\sigma}_\alpha,\hat{\sigma}_\beta \} =
2\delta_{\alpha,\beta}\hat{\mathbf{1}}$. However, this is not the case
for $S > 1/2$. The general anticommutator relations for
the spin operator $\hat{S}_\alpha$ are given by:
$\{\hat{S}_\alpha,\hat{S}_\beta\} = 4/N\delta_{\alpha\beta}
\hat{\mathbf{1}} + 2 g_{\alpha\beta\gamma} 
\hat{S}_\gamma$, where $g_{\alpha\beta\gamma}$ is the completely
symmetric tensor of the Lie algebra $su(N)$, and $N$ the number
of quantum level \cite{kimuraPLA03}. For $S = 1/2$ we have $N = 2$,
and $g_{\alpha\beta\gamma} = 0$. 

After working out the commutators and anticommutators we find with $\gamma =
g\mu_B/\hbar$: 
\begin{eqnarray}\label{MZTimeIII}
\frac{\mathrm{d}m_z}{\mathrm{d} t} &=& \gamma [m_xB_y-m_yB_x] -
\alpha_{\mathrm{l}} \gamma (\mathbf{B}\cdot\mathbf{m})\, m_z 
\nonumber \\ 
&-& \alpha_{\mathrm{tr}} \gamma [ m_z(\mathbf{B}\cdot \mathbf{m}) -
  B_z\underbrace{(\tilde{\mathbf{m}}\cdot\tilde{\mathbf{m}})}_{=1}
  [\langle \psi | \psi \rangle]^2] \nonumber  \;. \\
\end{eqnarray}  
Here, the definitions for $m_x = \langle \psi |
\hat{\sigma}_x |\psi \rangle$ and $m_y = \langle \psi |
\hat{\sigma}_y |\psi \rangle$ have been used and the assumption that
$\tilde{\mathbf{m}} = \langle \psi |\hat{\mathbf{\sigma}}| \psi
\rangle /\langle \psi |\psi \rangle$ is normalized:
$\tilde{\mathbf{m}} \cdot \tilde{\mathbf{m}} = 
\tilde{\mathbf{m}}^2 = 1$. 
   
The last equation can be written in a more compact form using the
vector triple product identity and 
$\mathbf{m} = \tilde{\mathbf{m}} \langle \psi |\psi \rangle$:
\begin{eqnarray}\label{MZTimeIV}
\frac{\mathrm{d}m_z}{\mathrm{d} t} &=& \gamma[\mathbf{m}
\times \mathbf{B}]_z - \alpha_{\mathrm{tr}} \gamma [\mathbf{m} \times
(\mathbf{m} \times \mathbf{B})]_z \nonumber \\ 
&-&\alpha_{\mathrm{l}} \gamma (\mathbf{B}\cdot\mathbf{m})\, m_z
\;. 
\end{eqnarray}  
The equations for $m_x$ and $m_y$ can be derived in a similar
way and therefore, we finally get:
\begin{eqnarray}\label{MZTimeV}
\frac{\mathrm{d}\mathbf{m}}{\mathrm{d} t} &=& \gamma\mathbf{m}
\times \mathbf{B} - \alpha_{\mathrm{tr}} \gamma \mathbf{m} \times
(\mathbf{m} \times \mathbf{B}) -
\alpha_{\mathrm{l}} \gamma (\mathbf{B}\cdot\mathbf{m})\, \mathbf{m}
\;. \nonumber \\ 
\end{eqnarray}    
This equation is identical to the classical Landau-Lifshitz-Bloch
equation, if we ignore the sign problem of the precessional motion
($\gamma \rightarrow -\gamma$).

\section{NUMERICAL EXAMPLES} \label{s:num}
In the last section \ref{s:proof} the proposal has been proved analytical for
the case of a single spin with $S = 1/2$. We have seen that in this
case the Schr\"odinger equation [Eq.~(\ref{TDSELLB2})] leads to an
equation for the expectation values $\mathbf{m} = \langle \psi
|\hat{\mathbf{S}} | \psi \rangle$ which is similar to the classical
Landau-Lifshitz-Bloch equation. To strengthen this statement we present
in this section computer simulations to show the correctness and the
possibilities of the given description.   

For the computer simulations we have solved the time dependent Schr\"odinger
equation [Eq.~(\ref{TDSELLB})] numerical for 
different scenarios. In the following we set $\hbar = 1$ which means
that the time will be in natural units: $t_{\mathrm{sim.}} = \hbar
t_{\mathrm{real}}$. Under the assumption that we have energies in
units of electronvolts the time scales of the simulations are in the
femtosecond regime.  

In the first scenario we assume a starting
configuration of one spin with spin quantum number $S = 1/2$ oriented
in $+x$-direction: $|\psi \rangle_{\mathrm{init}} = (|\!\uparrow\,\rangle +
|\!\downarrow\,\rangle)/\sqrt{2}$ and a length $|\langle
\hat{\mathbf{S}} \rangle|/\hbar S = 1$. Furthermore, an external field  
in $+z$-direction: $\mathbf{B} = B_z \hat{\mathbf{z}}$. The scenario
has been chosen in such a way that, following the description (point
4c), we can expect a behavior of the spin expectation value $\langle
\hat{\mathbf{S}} \rangle$ similar to the dynamics of a classical spin
$\mathbf{S}$ (except the different sense of rotation). The Hamiltonian
of this scenario is given by: 
\begin{equation} \label{HamSzene1}
\hat{\mathrm{H}} = - \frac{g\mu_B}{\hbar}B_z \hat{S}_z \;.
\end{equation}

Due to the relaxation terms the spin will relax into the direction of
the external field and shrink to the equilibrium length $|\langle
\hat{\mathbf{S}} \rangle_{\mathrm{eq}}|/\hbar S = 0.7$. 
For $\alpha_{\mathrm{tr}}$ and $\alpha_{\mathrm{l}}$ we use the
definitions:
\begin{eqnarray} \label{dampRob}
\alpha_{\mathrm{tr}} = \alpha^0_{\mathrm{tr}} \;\;\;\mathrm{and} \;\;\;
\alpha_{\mathrm{l}} =  \alpha^0_{\mathrm{l}} \cdot \left(\frac{|\langle
\hat{\mathbf{S}} \rangle| - |\langle \hat{\mathbf{S}} 
\rangle_{\mathrm{eq}}|}{\hbar S}\right) \;, 
\end{eqnarray}
with $\alpha^0_{\mathrm{tr}} = 0.02$ and $\alpha^0_{\mathrm{l}} = 0.04$.

\begin{figure}
\vspace{1mm} 
\includegraphics*[width=7.7cm,bb = 60 450 520 770]{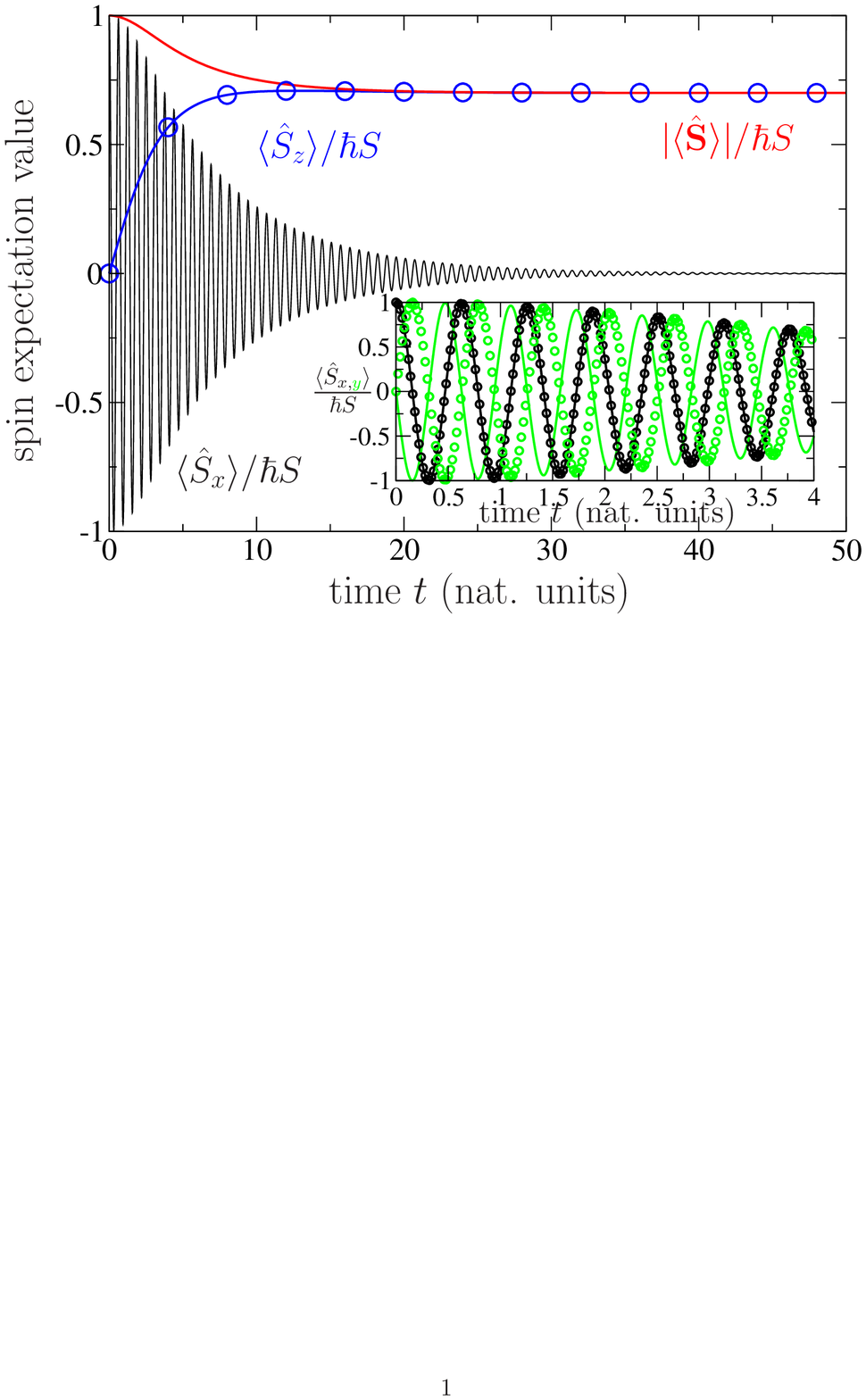}  
  \caption{(color online) Relaxation of a single spin ($S = 1/2$) with initial
    orientation in $+x$-direction and length $|\langle \hat{\mathbf{S}}
    \rangle|/ \hbar S = 1$ in an external field in
    $+z$-direction and a final spin length $|\langle \hat{\mathbf{S}}
    \rangle|/ \hbar S = 0.7$. The solid lines correspond to the quantum
    mechanical expectation values $\langle \hat{\mathbf{S}}
    \rangle$ and the circles the classical
    trajectories $\mathbf{S}$. 
    (Simulation parameter: $\hbar = 1$, $\gamma B_z = 10$,
    $\alpha_{\mathrm{tr}}^0 = 0.02$, $\alpha_{\mathrm{l}}^0 = 0.04$) 
}       
  \label{f:pic1}
\end{figure}

Fig.~\ref{f:pic1} presents the relaxation process of a
quantum spin ($S=1/2$) compared with a classical spin ($S=\infty$). In
the later case we have solved the Landau-Lifshitz equation. In both
cases the spin has been normalized and in the case of the classical
spin we have used the double field value: $2B_z$ 
instead of $B_z$. The doubling of the field value is necessary to make
the quantum mechanical and the classical trajectories comparable. The
reason for that are the different Zeeman energies qm:
$\hat{\mathrm{H}} = - g \mu_B 
\mathbf{B}\cdot\hat{\mathbf{\sigma}}/2$ [see Eq.~(\ref{HamHam})] and
cl.: $\mathrm{H} = - g \mu_B \mathbf{B}\cdot\mathbf{S}$.
Fig.~\ref{f:pic1} clearly shows that in this case we find a perfect
agreement between the classical trajectories $\mathbf{S}$ and quantum
mechanical expectation values $\langle \hat{\mathbf{S}} \rangle$. The
only difference appears for $\langle \hat{S}_x \rangle$. Here, we see
a phase difference of $180^\circ$ coming from the different rotation
senses of the precession in the classical and quantum spin dynamics
($\gamma \rightarrow -\gamma$ problem). However, the amplitude and
frequency are are same. 

To show the reliability of the given description we have performed
more complex simulations. The next scenario has an initial
configuration with a single spin with $S=1/2$ oriented in
$+z$-direction with $|\langle \hat{\mathbf{S}}  
\rangle|/\hbar S = 1$. As before we assume an external magnetic field in $+z$
direction plus an additional Gaussian field pulse:  
\begin{equation} \label{Gaussian}
\mathbf{B}_x(t) =
B_0^xe^{-\frac{1}{2}\left(\frac{t-t_0}{T_W}\right)^2}\hat{\mathbf{x}}  
\end{equation} 
in $x$-direction to excite the spin. 
Therefore, the Hamilton operator of this scenario is:
\begin{equation} \label{HamSzene2}
\hat{\mathrm{H}} = - \frac{g\mu_B}{\hbar}\left(B_x(t) \hat{S}_x + B_z
\hat{S}_z \right)\;,  
\end{equation}
with $B_x(t)$ given by Eq.~(\ref{Gaussian}).

For the damping parameters
$\alpha_{\mathrm{tr}}$ and $\alpha_{\mathrm{l}}$ the definitions given
by L.~Xu and S.~Zhang \cite{xuPE12,xuJAP13} have been used. Within
their publication L.~Xu and S.~Zhang have proposed the following von
Neumann equation:  
\begin{equation} \label{scBloch}
\frac{\mathrm{d}\hat{\rho}}{\mathrm{d}t} =
\frac{i}{\hbar}\left[\hat{\rho},\hat{\mathrm{H}}\right] -
\frac{\hat{\rho} - \hat{\rho}_{\mathrm{eq}}}{\tau_S} \;,
\end{equation}
which becomes with $\hat{\mathrm{H}} =
-\frac{g \mu_B}{\hbar} 
\mathbf{B}\cdot \hat{\mathbf{S}}$ in the classical limit:
\begin{equation} \label{scBloch_m}
\frac{\mathrm{d}\mathbf{m}}{\mathrm{d}t} = \gamma \mathbf{m} \times
\mathbf{B} - \frac{\mathbf{m} - \mathbf{m}_{\mathrm{eq}}}{\tau_S} \;.
\end{equation}
$\mathbf{m}_{\mathrm{eq}}$ is the equilibrium magnetization and
$\tau_S$ is a constant describing the strength of the relaxation.  

L.~Xu and S.~Zhang have shown that with the identity: 
$\mathbf{B} =
m^{-2}\left[\left(\mathbf{m}\cdot\mathbf{B}\right)\mathbf{m} -
  \mathbf{m} \times \left(\mathbf{m} \times
  \mathbf{B}\right)\right]$ 
this equation becomes:
\begin{equation} \label{clLLB}
\frac{\mathrm{d}\mathbf{m}}{\mathrm{d}t} = 
\gamma \mathbf{m} \times
\mathbf{B}  - \gamma \alpha_{\mathrm{tr}} \mathbf{m}  \times \left(\mathbf{m}
 \times \mathbf{B}\right) -  \gamma \alpha_{\mathrm{l}}
 \left(\mathbf{m} \cdot \mathbf{B} \right) \mathbf{m} \;,    
\end{equation} 
which is identical to the Landau-Lifshitz-Bloch Eq.~(\ref{LLB}). 

The damping parameters are given by: 
\begin{equation} \label{alpha_tr}
\alpha_{\mathrm{tr}} = \frac{m_{\mathrm{eq}}}{\gamma \tau_s m B} \;,
\end{equation}
as well as
\begin{equation} \label{alpha_l}
\alpha_{\mathrm{l}} = \frac{1}{\gamma
  \tau_s}\left[\frac{m}{\mathbf{m}\cdot\mathbf{B}} -
  \frac{m_{\mathrm{eq}}}{mB}\right]  \;.
\end{equation}
To make Eq.~(\ref{clLLB}) more general we replace $\tau_s$ in
Eq.~(\ref{alpha_tr}) by $\tau_{\mathrm{tr}}$ and in
Eq.~(\ref{alpha_l}) by $\tau_{\mathrm{l}}$. $\tau_{\mathrm{tr}}$ and
$\tau_{\mathrm{l}}$ are similar to $\tau_s$ constants. Furthermore, we 
replace the classical $\mathbf{m}$, $m = |\mathbf{m}|$, and
$m_{\mathrm{eq}}$ by their quantum mechanical counterparts $\langle
\hat{\mathbf{S}} \rangle$, $|\langle \hat{\mathbf{S}} \rangle|$, and
$|\langle \hat{\mathbf{S}} \rangle_{\mathrm{eq}}|$. The results are
the following two damping parameters:
\begin{equation} \label{alpha_tr_qm}
\alpha_{\mathrm{tr}} = \frac{\hbar}{g \mu_B \tau_{\mathrm{tr}}} \frac{|\langle
  \hat{\mathbf{S}} 
  \rangle_{\mathrm{eq}}|}{|\langle
  \hat{\mathbf{S}} \rangle| B} \;, 
\end{equation}
and
\begin{equation} \label{alpha_l_qm}
\alpha_{\mathrm{l}} = \frac{\hbar}{g \mu_B
  \tau_{\mathrm{l}}}\left[\frac{|\langle \hat{\mathbf{S}} \rangle|}{\langle
    \hat{\mathbf{S}} \rangle\cdot\mathbf{B}} - 
  \frac{|\langle \hat{\mathbf{S}} \rangle_{\mathrm{eq}}|}{|\langle
    \hat{\mathbf{S}} \rangle| B}\right]  \;. 
\end{equation}

\begin{figure}
\vspace{1mm} 
\includegraphics*[width=7.7cm,bb = 60 450 520 770]{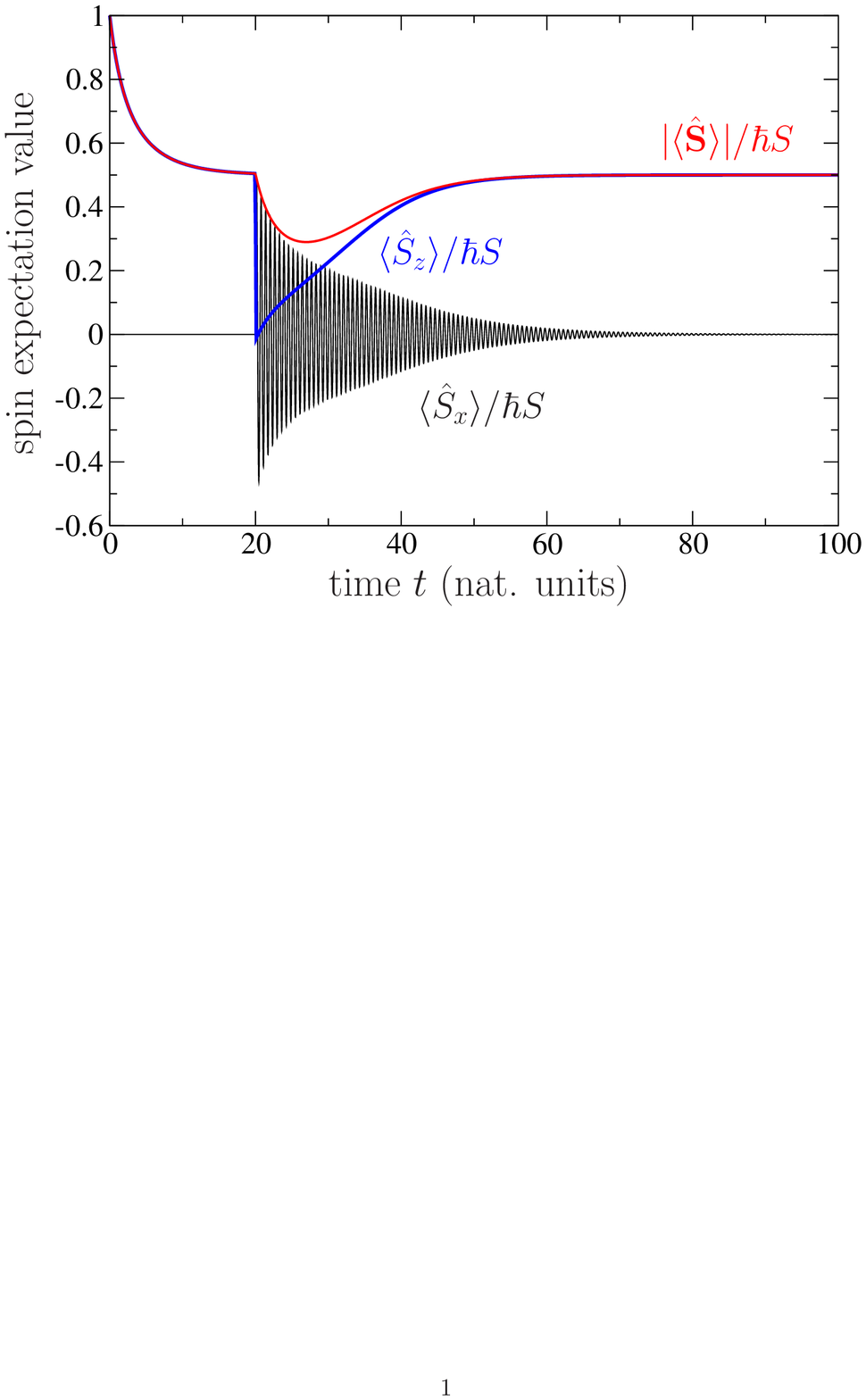}  
  \caption{(color online) Longitudinal relaxation and combined transversal +
    longitudinal relaxation after a Gaussian field pulse at $t = 20$
    for a quantum spin with $S = 1/2$, initial and final spin length $|\langle
  \hat{\mathbf{S}} \rangle_{\mathrm{init}}|/ \hbar S = 1$ respectively $|\langle
  \hat{\mathbf{S}} \rangle_{\mathrm{final}}|/ \hbar S =
  0.5$. (Simulation parameter: $\hbar = 1$, $\gamma B_z = 10$, $\gamma B_0^x =
  46.54$, $(\gamma \tau_{\mathrm{tr}})^{-1} = 0.1$, $(\gamma
  \tau_{\mathrm{l}})^{-1} = 0.2$, $t_0 = 20$, $T_W = 0.02$)}        
  \label{f:pic2}
\end{figure}

Fig. \ref{f:pic2} shows the $x$ and $z$ components of the spin
expectation value $\langle \hat{\mathbf{S}} \rangle$ as well as the
length $|\langle \hat{\mathbf{S}} \rangle|$ of a single spin in an
external field oriented in $+z$ direction as function of time. The
initial spin is oriented parallel to the external field and has a
length of $|\langle \hat{\mathbf{S}} \rangle| = 1$. The equilibrium
length $|\langle \hat{\mathbf{S}} \rangle_{\mathrm{eq}}|$ has been
chosen as $|\langle \hat{\mathbf{S}} \rangle_{\mathrm{eq}}| = 0.5$.
Therefore, and due to the fact that there is only the external field
in $+z$ direction only the longitudinal relaxation contributes to the
dynamics. Fig. \ref{f:pic2} clearly shows that the $z$ component of
$\langle \hat{\mathbf{S}} \rangle$ decays exponentially with the time
until it reaches the equilibrium length $|\langle \hat{\mathbf{S}}
\rangle_{\mathrm{eq}}| = 0.5$. After reaching the equilibrium a
Gaussian field pulse has been applied bringing the $z$ component
$\langle \hat{S}_z \rangle$ close to zero. After the field pulse the
spin relaxes back to equilibrium, but this time all three terms:
precession, transverse and longitudinal relaxation contribute to the
dynamics.
  
The last example shall demonstrate that the given description is not
restricted to a single spin. In the following we assume two spins $S =
1/2$ antiferromagnetically exchange coupled and where the first spin
can be manipulated by an external field:
\begin{equation}
\hat{\mathrm{H}} = J\frac{\hat{\mathbf{S}}_1 \cdot
\hat{\mathbf{S}}_2}{\hbar^2}  - \frac{g\mu_B}{\hbar} B_1^z(t) \hat{S}_1^z \;
\end{equation}
The first term describes the antiferromagnetic exchange coupling with
$J >0$. The second term describes the coupling between first spin and
an external field which is time dependent. This external field can be
seen as a rough description of an electric current of an
spin-polarized scanning tunneling microscope \cite{stapelfeldtPRL11}
or as an approximate description of the coupling to a magnetic island
as described in \cite{khajetooriansSCIENCE11}. In both 
cases we assume that we can switch the field and therefore the
influence on and off.  
\begin{figure}
\vspace{1mm} 
\includegraphics*[width=7.7cm,bb = 60 450 520 770]{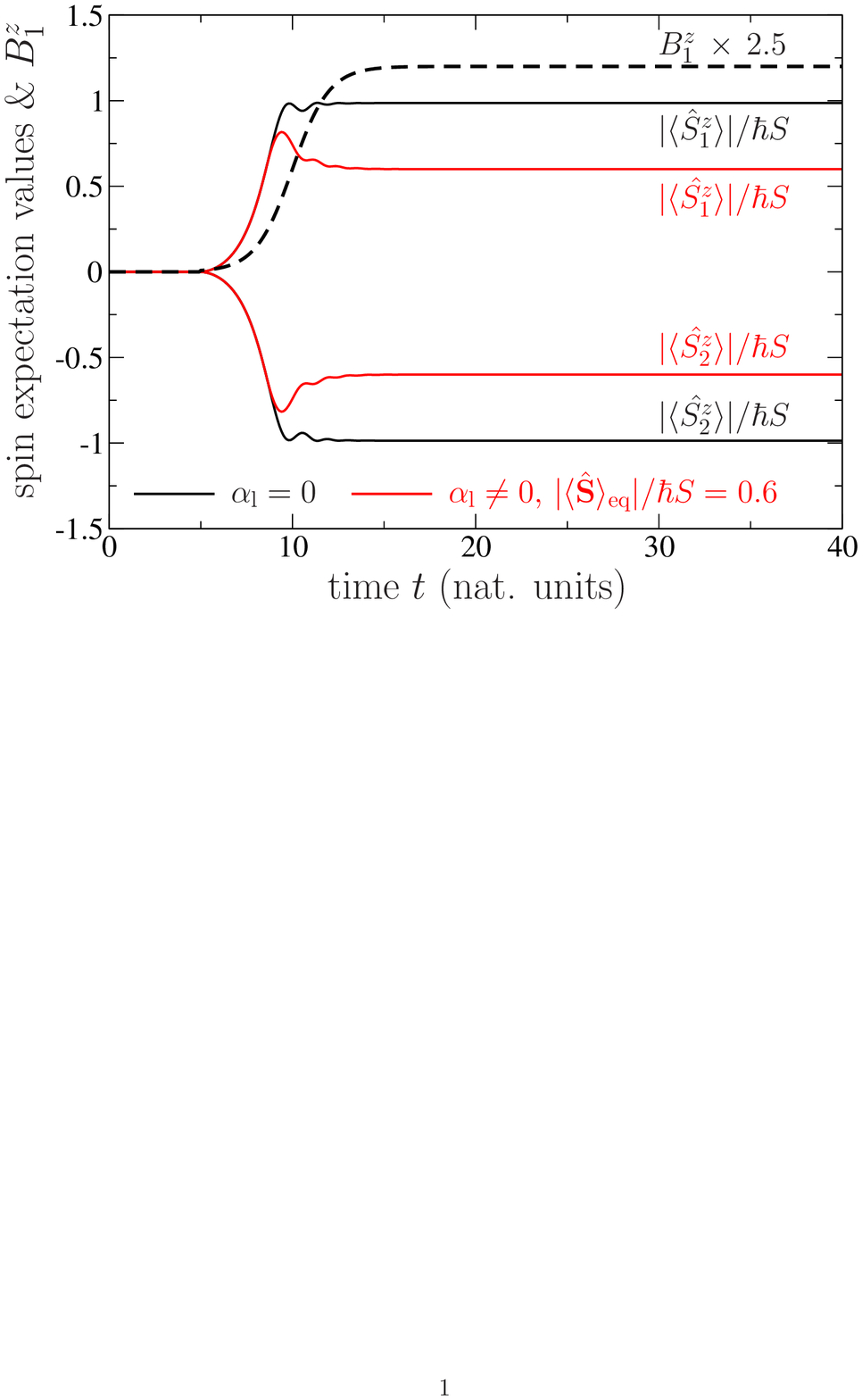}  
  \caption{(color online) Normalized spin expectation values $\langle
    \hat{S}_n^z \rangle / \hbar S$, $n \in \{1,2\}$ and external field $B_1^z$
    ($2.5$ times smaller than in real) as function of time $t$. The different
    colors correspond to the different assumptions of
    $\alpha_{\mathrm{l}}$: $\alpha^0_{\mathrm{l}} = 0$  and
    $\alpha^0_{\mathrm{l}} = 1.0$ together with 
    $|\langle\hat{\mathbf{S}}\rangle_{\mathrm{eq}}|/\hbar S =
    0.6$. (Simulation parameter: $\hbar = 1$, $J/\hbar^2 = 1$, $\gamma
    B_0^z = 3$, $\kappa = 0.5$, $t_0 = 10$, $\alpha_{\mathrm{tr}}^0 = 0.2$,
    $\alpha_{\mathrm{l}}^0 = 1.0$)}          
  \label{f:pic3}
\end{figure}

In the case of a magnetic island it means that we e.g. increase the
temperature above the Curie temperature $T_C$ to switch the field off
and let the island cool down to switch it on again. We further assume
that we start with a zero external field which increases with the time:
\begin{equation}
B_1^z(t) = B_0^z \mathrm{tanh}[\kappa (t-t_0)] + B_0^z \;,
\end{equation} 
where $B_0^z = 3.0$ and $\kappa = 0.5$ are constants describing the
maximum field strength and the inclination with the time and $t_0 = 10$. 

As long as the field is switched off ($B_1^z(t) = 0$) the spins are in
the ground state configuration which means in this case the singlet state:
\begin{equation} \label{Singulett}
|\psi\rangle = \frac{1}{\sqrt{2}} \left(|\uparrow \downarrow \,\rangle -
|\downarrow \uparrow \,\rangle \right) \;.
\end{equation}
With $B_1^z(t) > 0$ the first spin becomes stabilized and we find as
final state:
\begin{equation} \label{Neelstate}
|\psi\rangle =
\sqrt{|\langle\hat{\mathbf{S}}\rangle_{\mathrm{eq}}|} |\uparrow
\downarrow \,\rangle \;.  
\end{equation}
$|\langle\hat{\mathbf{S}}\rangle_{\mathrm{eq}}|$ has been
determined at the beginning of the calculation as part of the
definition of $\alpha_{\mathrm{l}}$ [see Eq.~(\ref{dampRob}) or
(\ref{alpha_l_qm})]. During the simulation for the damping terms the
definition Eq.~(\ref{dampRob}) with $\alpha^0_{\mathrm{tr}} = 0.2$ and
$\alpha^0_{\mathrm{l}} = 1.0$ has been used. Fig.~\ref{f:pic3} shows
the normalized spin expectation 
values $\langle \hat{S}_n^z \rangle / \hbar S$, $n \in \{1,2\}$ of the first
and second spin, as well as the time dependent external field
$B_1^z(t)$. The zero expectation values $\langle \hat{S}_1^z \rangle /
\hbar S = \langle \hat{S}_2^z \rangle / \hbar S = 0$ correspond to the
initial singulet state [Eq.~(\ref{Singulett})] and $B_1^z = 0$. With
$B_1^z > 0$ the expectation values $\langle \hat{S}_n^z \rangle /
\hbar S$ increase / decrease to their final values $\pm
|\langle\hat{\mathbf{S}}\rangle_{\mathrm{eq}}|$ corresponding to
$|\psi \rangle_{\mathrm{final}}$ given by Eq.~(\ref{Neelstate})
(classical N\'eel state). The other expectation values $\langle
\hat{S}_n^x \rangle / \hbar S$ and $\langle \hat{S}_n^y \rangle /
\hbar S$ are zero for all times. 

\section{SUMMARY} \label{s:summary}
Within this publication the way how to derive the time dependent
Schr\"odinger equation which can be seen as the quantum mechanical
analog to the classical Landau-Lifshitz-Bloch equation has been
demonstrated. The starting point is the Landau-Lifshitz-Bloch equation
itself. From this equation the corresponding von Neumann equation can
be deduced. And the von Neumann equation is the starting point to
derive the corresponding time dependent Schr\"odinger
equation. Therefore, with von Neumann equation and the time dependent
Schr\"odinger equation we have two equations which can be seen as
quantum mechanical analogs to the classical Landau-Lifshitz-Bloch
equation. This opens the opportunity to extend the spin dynamics with
transversal and longitudinal relaxation to the quantum regime and to
compare the classical with the quantum spin dynamics.

The correctness of the time dependent Schr\"odinger equation and
therefore indirectly also of the von Neumann equation has been proved
analytical and with computer simulations. It has been shown that derived
Schr\"odinger equation can lead to the same dynamics as the classical
Landau-Lifshitz-Bloch equation. However, in the most scenarios we have
to expect a different dynamics due to quantum effects. With the given
description we have a general description of the quantum spin dynamics
with transversal and longitudinal relaxation which is simple to
understand. However, the given description will not give an insight to the 
microscopic processes of the damping. The energy gain and loss is
introduced by phenomenological damping functions. This is the same in
the case of the classical description. This can be seen as a
disadvantage. On the other hand the Landau-Lifshitz-Gilbert as well as
the Bloch equation became successful due to their simplicities and the
fact that it is not necessary to know the underlying damping
mechanism. In the given description this is the same and can be seen
as an advantage. Furthermore, due to the fact that the damping
parameters are no longer constant, but functions the given description
is quite general. E.g. using the description of coherent states for
$|\psi \rangle$ \cite{schliemannJPCONDMAT97,balakrishnanJPCondMat90}
together with the definition for $\alpha_{\mathrm{tr}}$ and
$\alpha_{\mathrm{l}}$ in \cite{nievesPRB14} reproduces the
Landau-Lifshitz-Bloch equation which has been successful used to
describe ultrafast magnetization dynamics measurements.


\bibliography{Cite}

\end{document}


\date{\today}

\title{Supplementary Material: Derivation of a time dependent
  Schr\"odinger equation as quantum mechanical Landau-Lifshitz-Bloch equation}  

\author{R.\ Wieser}
\affiliation{1. International Center for Quantum Materials, Peking
  University, Beijing 100871, China \\  
2. Collaborative Innovation Center of Quantum Matter, Beijing 100871, China}

\pacs{75.78.-n, 75.10.Jm, 75.10.Hk}
\maketitle

\section{Introduction} \label{s:Intro}
The following sections gives an overview about the quantum spin
dynamics using the von Neumann equation with additional transversal
and longitudinal relaxation terms as equation of motion to describe
quantum spin dynamics. The work is still in progress and maybe not all
the derivations and descriptions are not hundred percent correct. However, the
idea of this supplementary is to give an introduction in this topic and
to present some concepts.  

\section{Derivation of the von Neumann Eq. which can be seen
as the quantum mechanical analog to the Landau-Lifshitz
equation} \label{s:Liouville}  

Starting point is the non-Hermitian Hamiltonian operator
\begin{equation}
\hat{\cal H} = \hat{\mathrm{H}} - i \alpha \hat{\mathrm{H}} \;,
\end{equation}
with $\hat{\mathrm{H}}$ a Hermitian Heisenberg model Hamilton operator and
$\alpha \in {\rm I\!R}^+_0$ a constant. Such an Hamiltonian leads
to energy dissipation and at the same time does not conserve the norm of
the wave function: 
\begin{equation}
n = \langle \psi(t) |\psi(t) \rangle = \langle \psi_0 |
e^{-2\alpha\hat{\mathrm{H}}t} |\psi_0 \rangle = e^{-2\alpha\langle
    \hat{\mathrm{H}} \rangle t} \;.   
\end{equation}
However, the norm can be conserved by replacing $i \alpha
\hat{\mathrm{H}}$ by $i \alpha(\hat{\mathrm{H}} - \langle
\hat{\mathrm{H}} \rangle)$  
\begin{equation} \label{PsiTime}
|\psi (t) \rangle = e^{-i\hat{\mathrm H}t}
e^{-\alpha\hat{\mathrm{H}}t}e^{\alpha \langle \hat{\mathrm{H}} \rangle t} |\psi_0 
\rangle \;.
\end{equation}
In this case the norm keeps constant $n = 1$. 

The corresponding Schr\"odinger equation is given by: 
\begin{equation} \label{Gisin}
\mathrm{i} \hbar \frac{\mathrm{d}}{\mathrm{d}t}|\psi(t)\rangle = 
(\hat{\mathrm{H}} -\mathrm{i}\alpha [\hat{\mathrm{H}} - 
\langle \hat{\mathrm{H}} \rangle])| \psi(t)\rangle \;.
\end{equation}

Eq.~(\ref{Gisin}) can be easily written as:
\begin{equation} \label{GisinII}
i \hbar \frac{\mathrm{d}}{\mathrm{d}t} |\psi \rangle =
\Big(\hat{\mathrm{H}} - i \alpha \Big[\hat{\mathrm{H}}, |\psi \rangle
\langle \psi |\Big]\Big)|\psi \rangle \;.
\end{equation}
The corresponding conjugate transposed equation is given by:
\begin{equation} \label{GisinIII}
-i \hbar \frac{\mathrm{d}}{\mathrm{d}t} \langle \psi | =  \langle \psi | \left(
\hat{\mathrm{H}} + i \alpha \Big[|\psi \rangle
\langle \psi |, \hat{\mathrm{H}}\Big] \right) \;.
\end{equation}
With these equations, we are able to construct a von Neumann
equation: 
\begin{eqnarray} 
\frac{\mathrm{d}}{\mathrm{d}t} \Big(|\psi \rangle \langle \psi |\Big)
&=& \frac {\mathrm{d} |\psi \rangle}{\mathrm{d}t} \langle \psi | +
 |\psi \rangle \frac{\mathrm{d} \langle \psi |}{\mathrm{d}t} \nonumber \\
&=& -\frac{i}{\hbar}\Big(\hat{\mathrm{H}} - i \alpha
\Big[\hat{\mathrm{H}}, |\psi \rangle \langle \psi |\Big]\Big)|\psi
\rangle \langle \psi |+ \frac{i}{\hbar}|\psi \rangle   \langle \psi |
\Big(\hat{\mathrm{H}} + i \alpha \Big[|\psi \rangle
\langle \psi |, \hat{\mathrm{H}} \Big] \Big) \nonumber \\
&=& \frac{i}{\hbar} \Big[|\psi \rangle   \langle \psi |,
\hat{\mathrm{H}} \Big] -  \frac{\alpha}{\hbar} \Big[|\psi \rangle
\langle \psi |,\Big[|\psi \rangle \langle \psi |, \hat{\mathrm{H}} \Big]
\Big] \nonumber \;,
\end{eqnarray}
and finally with $\hat{\rho} = |\psi \rangle \langle \psi |$  
\begin{equation}\label{vNE}
\frac{\mathrm{d}\hat{\rho}}{\mathrm{d}t} = \frac{i}{\hbar} [\hat{\rho},
\hat{\mathrm{H}} ] - \frac{\alpha}{\hbar} [\hat{\rho},[\hat{\rho},
\hat{\mathrm{H}} ]]\;. 
\end{equation}

\section{Derivation of the Landau-Lifshitz-Bloch Eq. starting from the
von Neumann Eq. using geometric Algebra} \label{s:qLLB}

Starting point is the following von Neumann equation:
\begin{equation} \label{LiouvilleSecII}
\frac{\mathrm{d}\hat{\rho}}{\mathrm{d}t} =
\frac{i}{\hbar}[\hat{\rho},\hat{\mathrm{H}}] -
  \frac{\alpha_{\mathrm{tr}}}{\hbar}[\hat{\rho},[\hat{\rho},\hat{\mathrm{H}}]]
  + 2\frac{\alpha_{\mathrm{l}}}{\hbar}
  (\hat{\rho}\hat{\mathrm{H}})\hat{\rho} \;,  
\end{equation} 
with the assumption that the density operator $\hat{\rho}$ is given by:
\begin{eqnarray} \label{Rho}
\hat{\rho} &=& \frac{1}{2}
\left(\hat{\mathbf{1}} + \langle \hat{\bold{\sigma}} \rangle \cdot
\hat{\bold{\sigma}} \right) \;,  
\end{eqnarray}
Furthermore, we assume that the Hamiltonian can be written as:
\begin{equation} \label{HamHam}
\hat{\mathrm{H}} = - \frac{g\mu_B}{2} \mathbf{B} \cdot \hat{\bold{\sigma}} \;,
\end{equation}
with $g$ the g-factor and $\mu_B$ Bohr magneton for the correct dimension.
$\hat{\bold{\sigma}} = (\hat{\sigma}_x,\hat{\sigma}_y,\hat{\sigma}_z)$
is the Pauli vector with $\hat{\sigma}_\eta$, $\eta \in \{x,y,z\}$ the
Pauli Matrices and $\hat{\mathbf{1}}$ the identity matrix:
\begin{eqnarray}
\hat{\sigma}_x = \left(\begin{array}{cc} 0 & 1 \\ 1 &
  0 \end{array}\right) \;,\;\;\;\hat{\sigma}_y =
\left(\begin{array}{cc} 0 & -i \\ i & 
  0 \end{array}\right) \;,\;\;\;\hat{\sigma}_z =
\left(\begin{array}{cc} 1 & 0 \\ 0 & 
  -1 \end{array}\right) \;,\;\;\;
\hat{\mathbf{1}} = \left(\begin{array}{cc} 1 & 0 \\ 0 &
  1 \end{array}\right)
\;.
\end{eqnarray}
A view words about Eq.~(\ref{Rho}) and (\ref{HamHam}): As in the
Euclidean space where we are able to express any vector in 
${\rm I\!R}^3$ as a linear combination of the basis vectors $\hat{\mathbf{x}}$,
$\hat{\mathbf{y}}$, and $\hat{\mathbf{z}}$: $\mathbf{r} =
x\hat{\mathbf{x}} + y\hat{\mathbf{y}} + z\hat{\mathbf{z}}$, we can (as
an isomorphism) do the same with the Pauli matrices e.g.:
\begin{eqnarray}
\cal{P} &=& \mathbf{P} \cdot \hat{\bold{\sigma}} = 
P_x\hat{\sigma}_x + P_z\hat{\sigma}_z + P_z\hat{\sigma}_z \,. 
\end{eqnarray}
Here, we have set $\mathbf{P} = \langle \hat{\bold{\sigma}} \rangle$.
Or, alternatively written as matrix:
\begin{eqnarray}
{\cal P} = \left(\begin{array}{cc} P_z & P_x - i P_y \\ P_x + i P_y &
  -P_z \end{array}\right) \;.
\end{eqnarray}
${\cal B} = -\mathbf{B} \cdot \hat{\bold{\sigma}}$ is defined in a
similar way.

Lets focus on the von Neumann equation. With aid of (\ref{Rho}) the
left hand side of Eq.~(\ref{LiouvilleSecII}) can be written  as: 
\begin{eqnarray}
\frac{\mathrm{d}\hat{\rho}}{\mathrm{d}t} =
\frac{\mathrm{d}\mathbf{P}}{\mathrm{d}t}
\cdot \frac{\hat{\bold{\sigma}}}{2} \;.
\end{eqnarray}
Please notice, here we are in the Schr\"odinger picture, meaning that
the operators are time independent. 

The precessional term (first term) of Eq.~(\ref{LiouvilleSecII}) contains
the following commutator:
\begin{eqnarray} \label{innerCom}
[\hat{\rho},\hat{\mathrm{H}}] = \frac{g \mu_B}{4}[{\cal P},{\cal B}] =
\frac{g \mu_B}{2} {\cal P} \wedge {\cal B} \;.
\end{eqnarray}
The wedge product in Eq.~(\ref{innerCom}) is defined as:
\begin{eqnarray}
{\cal P} \wedge {\cal B} = i P^n B^m \epsilon_{nml}
\hat{\sigma}_l \;,
\end{eqnarray}
where $\epsilon_{nml}$ is the Levi-Civita tensor. Here, the Einstein
sum convention has been used. Comparison with the vector product:
\begin{eqnarray}
\mathbf{a} \times \mathbf{b} = a^nb^m \epsilon_{nml}
\hat{\mathbf{e}}_l \;,
\end{eqnarray}
where $\hat{\mathbf{e}}_l$ is a unit vector perpendicular to
$\mathbf{a}$ and $\mathbf{b}$, leads to
\begin{eqnarray} 
{\cal P} \wedge {\cal B} = i(\mathbf{P} \times
\mathbf{B})^l \hat{\sigma}_l = i(\mathbf{P} \times \mathbf{B})\cdot
\hat{\bold{\sigma}} \;. 
\end{eqnarray}

Next term is the transversal relaxation term which contains the double
commutator: 
\begin{eqnarray}
[\hat{\rho},[\hat{\rho},\hat{\mathrm{H}}]] = \frac{g \mu_B}{8}[{\cal P},[\cal{
      P},{\cal B}]]  \;. 
\end{eqnarray}
The inner commutator has been already calculated [see Eq.~(\ref{innerCom}) and
  following equations]. The result can be written as:
\begin{eqnarray}
[\hat{\rho},\hat{\mathrm{H}}]  = \frac{g \mu_B}{4}[{\cal P},{\cal B}] =
\frac{ig \mu_B}{2} (\mathbf{P} \times \mathbf{B})\cdot 
\hat{\bold{\sigma}} = \frac{ig \mu_B}{2} \mathbf{A} \cdot
\hat{\bold{\sigma}} = \frac{ig \mu_B}{2} {\cal A}\;,
\end{eqnarray}
therefore:
\begin{eqnarray}
[\hat{\rho},[\hat{\rho},\hat{\mathrm{H}}]] = \frac{ig \mu_B}{4}[{\cal P},{\cal
    A}] = \frac{i^2g \mu_B}{2} (\mathbf{P} \times \mathbf{A})\cdot
\hat{\bold{\sigma}} = -g \mu_B(\mathbf{P} \times (\mathbf{P} \times
\mathbf{B}))\cdot \frac{\hat{\bold{\sigma}}}{2} \;.  
\end{eqnarray}
With this results and the gyromagnetic ratio $\gamma = g \mu_B / \hbar$ ,
the von Neumann equation without longitudinal relaxation term: 
\begin{equation} \label{Liouville2}
\frac{\mathrm{d}\hat{\rho}}{\mathrm{d}t} =
\frac{i}{\hbar}[\hat{\rho},\hat{\mathrm{H}}] -
  \frac{\alpha_{\mathrm{tr}}}{\hbar}[\hat{\rho},[\hat{\rho},\hat{\mathrm{H}}]]
  \;,   
\end{equation} 
is given as:
\begin{equation} \label{Liouville3}
\frac{\mathrm{d}\mathbf{P}}{\mathrm{d}t} = \gamma\mathbf{P}
 \times \mathbf{B} - \alpha_{\mathrm{tr}}\gamma(\mathbf{P}
 \times (\mathbf{P} \times \mathbf{B})) \;.   
\end{equation} 
Here, the $\cdot \hat{\bold{\sigma}}/2$ term on both sides has been
already skipped. 

The results above can be proved by the direct multiplication
and subtraction of the matrices. This shall be shown only for the
precessional term, the transversal relaxation term is similar. For the
commutator $[\hat{\rho},\hat{\mathrm{H}}] = -\frac{\gamma}{2}[{\cal
    P},{\cal B}] = -\frac{\gamma}{4}(\cal{P}{\cal B} - {\cal B}{\cal
  P})$ we have: 
\begin{eqnarray} \label{matrix1}
{\cal P}{\cal B} = \left(\begin{array}{cc} P_zB_z + (P_x - iP_y)(B_x +
  iB_y) \;&\; P_z(B_x - iB_y) - B_z(P_x - iP_y) \\ B_z(P_x + iP_y) -
  P_z(B_x + iB_y) \;&\; P_zB_z + (P_x + iP_y)(B_x - iB_y) \end{array}
\right) 
\end{eqnarray}
and
\begin{eqnarray} \label{matrix2}
{\cal B}{\cal P} = \left(\begin{array}{cc} B_zP_z + (B_x - iB_y)(P_x +
  iP_y) \;&\; B_z(P_x - iP_y) - P_z(B_x - iB_y) \\ P_z(B_x + iB_y) -
  B_z(P_x + iP_y) \;&\; B_zP_z + (B_x + iB_y)(P_x - iP_y) \end{array}
\right) 
\end{eqnarray} 
and therefore:
\begin{eqnarray} \label{matrix3}
{\cal P}{\cal B}-{\cal B}{\cal P} &=& 2i \left(\begin{array}{cc} (P_xB_y -
  P_yB_x) \;&\; (P_yB_z - P_zB_y) -i(P_zB_x - P_xB_z) \\
  (P_yB_z - P_zB_y) +i(P_zB_x - P_xB_z) \;&\; -(P_xB_y -
  P_yB_x) \end{array} \right) \nonumber \\ 
&=& 2i \left(\begin{array}{cc} (\mathbf{P} \times \mathbf{B})_z \;&\;
  (\mathbf{P} \times \mathbf{B})_x -i(\mathbf{P} \times \mathbf{B})_y
  \\ (\mathbf{P} \times \mathbf{B})_x +i(\mathbf{P} \times
  \mathbf{B})_y \;&\; -(\mathbf{P} \times \mathbf{B})_z \end{array}
\right) \;. 
\end{eqnarray} 
Finally with:
\begin{equation}
[{\cal P},{\cal B}] = {\cal P}{\cal B}-{\cal B}{\cal P} = 2i
(\mathbf{P} \times \mathbf{B}) \cdot \hat{\bold{\sigma}} \;,  
\end{equation}
we find:
\begin{equation}
\frac{i}{\hbar}[\hat{\rho},\hat{\mathrm{H}}] = -\frac{i\gamma}{4}[{\cal
    P},{\cal B}] = \gamma
(\mathbf{P} \times \mathbf{B}) \cdot \frac{\hat{\bold{\sigma}}}{2} \;.
\end{equation}

In the case of the longitudinal relaxation we have the following term:
\begin{eqnarray}
\frac{2\alpha_{\mathrm{l}}}{\hbar}(\hat{\rho}\hat{\mathrm{H}})\hat{\rho}
= -\frac{2\alpha_{\mathrm{l}}\gamma}{4 \hbar} \Big((\mathbf{P}\cdot
\hat{\bold{\sigma}}) (\mathbf{B} \cdot \hat{\bold{\sigma}})\Big)
 \left(\mathbf{P}\cdot\frac{\hat{\bold{\sigma}}}{2}\right)  =
-\frac{\alpha_{\mathrm{l}}\gamma}{2} \langle {\cal P}, {\cal B} \rangle
\cdot \left(\mathbf{P}\cdot\frac{\hat{\bold{\sigma}}}{2}\right) \;. 
\end{eqnarray}
As in the classical LLB equation where we have with
$(\mathbf{m}\cdot\mathbf{B})\mathbf{m}$ a scalar product (inner
product of two vectors in the Euclidian space) we have to deal here
with the inner product between the matrices ${\cal P}$ and 
${\cal B}$: 
\begin{eqnarray}
\langle {\cal P},{\cal B} \rangle = \mathrm{Tr}({\cal P}{\cal B}^\star) \;.
\end{eqnarray}
The star index ``$\star$'' means conjugate transposed. In our case,
${\cal B}$ is Hermitian ${\cal B}^\star = {\cal B}$ and therefore:
\begin{eqnarray}
\langle {\cal P},{\cal B} \rangle = \mathrm{Tr}({\cal P}{\cal B}) \;.
\end{eqnarray}
With: 
 \begin{eqnarray}
\mathrm{Tr}\left({\cal P}{\cal B}\right) &=&
\mathrm{Tr}\left(\begin{array}{cc} P_zB_z + (P_x - iP_y)(B_x + 
  iB_y) \;&\; P_z(B_x - iB_y) - B_z(P_x - iP_y) \\ B_z(P_x + iP_y) -
  P_z(B_x + iB_y) \;&\; P_zB_z + (P_x + iP_y)(B_x - iB_y) \end{array}
\right) = 2\mathbf{P}\cdot \mathbf{B} 
\nonumber \;,
\end{eqnarray}
we get finally:
\begin{eqnarray}
\langle {\cal P},{\cal B} \rangle = 2 \left(\mathbf{P}\cdot
\mathbf{B}\right) \;,   
\end{eqnarray}
where $\mathbf{P} \cdot \mathbf{B}$ is the scalar product between
two vectors in ${\rm I\!R}^3$ and therefore a scalar.

Using this result we are able to write:
\begin{eqnarray}
\frac{2\alpha_{\mathrm{l}}}{\hbar}(\hat{\rho}\hat{\mathrm{H}})\hat{\rho}
= -\frac{\alpha_{\mathrm{l}}\gamma}{2} \langle {\cal P},{\cal B} \rangle 
\cdot\left(\mathbf{P}\cdot\frac{\hat{\bold{\sigma}}}{2}\right)  =
-\alpha_{\mathrm{l}}\gamma (\mathbf{P} \cdot \mathbf{B})  
\cdot\left(\mathbf{P}\cdot\frac{\hat{\bold{\sigma}}}{2}\right)
 \;, 
\end{eqnarray}
and therefore the von Neumann Eq.~(\ref{LiouvilleSecII}) as:
\begin{equation} \label{Liouville4}
\frac{\mathrm{d}\mathbf{P}}{\mathrm{d}t} = \gamma \mathbf{P}
 \times \mathbf{B} - \alpha_{\mathrm{tr}}\gamma (\mathbf{P}
 \times (\mathbf{P} \times \mathbf{B}))
 -\alpha_{\mathrm{l}}\gamma (\mathbf{P} \cdot
 \mathbf{B}) \cdot \mathbf{P}\;.   
\end{equation} 
Again, the $\cdot \hat{\bold{\sigma}}/2$ term has been skipped on both
sides of this differential equation.

\section{Analytical proof using the Heisenberg equation} \label{s:Heisenberg}
To derive the Heisenberg equation:
\begin{eqnarray} \label{Heisenberg}
\frac{\mathrm{d}\langle\hat{\mathbf{S}}\rangle}{\mathrm{d}t} =
-\frac{i}{\hbar}\langle[\hat{\mathbf{S}},\hat{\mathrm{H}}]\rangle -
  \frac{\alpha_{\mathrm{tr}}}{\hbar}\langle\{\hat{\mathbf{S}},\hat{\mathrm{H}}\}
\rangle  + 2\frac{\alpha_{\mathrm{tr}} + \alpha_{\mathrm{l}} }{\hbar} \langle
  \hat{\mathrm{H}} \rangle \langle \hat{\mathbf{S}} \rangle 
  \;, 
\end{eqnarray} 
we can start from the von Neumann Eq.~(\ref{LiouvilleSecII}) and write this
equation in the alternative form:
\begin{equation}
[\hat{\rho},[\hat{\rho},\hat{\mathrm{H}}]] =
\{ \hat{\rho},\hat{\mathrm{H}} \} -
2\hat{\rho}\hat{\mathrm{H}}\hat{\rho} \;,
\end{equation} 
\begin{eqnarray} \label{LiouvilleAlternative}
\frac{\mathrm{d}\hat{\rho}}{\mathrm{d}t} =
\frac{i}{\hbar}[\hat{\rho},\hat{\mathrm{H}}] -
  \frac{\alpha_{\mathrm{tr}}}{\hbar} \{ \hat{\rho},\hat{\mathrm{H}}\}
  + 2\frac{\alpha_{\mathrm{tr}} + \alpha_{\mathrm{l}}}{\hbar}
  (\hat{\rho}\hat{\mathrm{H}})\hat{\rho} \;.
\end{eqnarray}
After adding $\mathrm{Tr}$ and $\hat{\mathbf{S}}$ to calculate the spin
expectation values $\langle\hat{\mathbf{S}}\rangle$ we end up with the
Heisenberg Eq.~(\ref{Heisenberg}). It is convenient to normalize the spin
expectation values $\langle\hat{\mathbf{S}}\rangle$ (dividing both sides of
Eq.~(\ref{Heisenberg}) by $\hbar S$). In the case of $S = 1/2$ this
leads to:
\begin{eqnarray} \label{Heisenberg2}
\frac{\mathrm{d}\langle\hat{\bold{\sigma}}\rangle}{\mathrm{d}t} =
-\frac{i}{\hbar}\langle[\hat{\bold{\sigma}},\hat{\mathrm{H}}]\rangle -
  \frac{\alpha_{\mathrm{tr}}}{\hbar}\langle\{\hat{\bold{\sigma}},
\hat{\mathrm{H}}\} \rangle  + 2\frac{\alpha_{\mathrm{tr}} +
  \alpha_{\mathrm{l}} }{\hbar} \langle 
  \hat{\mathrm{H}} \rangle \langle \hat{\bold{\sigma}} \rangle 
  \;. 
\end{eqnarray} 
At this point we have to say that we have to add an additional
$\langle \psi | \psi \rangle$ to the second term to be conform with
the time dependent Schr\"odinger equation given in the paper. We will
see that this term is also needed here. Then, with
$\tilde{\alpha}_{\mathrm{tr}} = \alpha_{\mathrm{tr}}\langle \psi |
\psi \rangle$ the final Heisenberg equation is given by:
\begin{eqnarray} \label{Heisenberg3}
\frac{\mathrm{d}\langle\hat{\bold{\sigma}}\rangle}{\mathrm{d}t} =
-\frac{i}{\hbar}\langle[\hat{\bold{\sigma}},\hat{\mathrm{H}}]\rangle -
  \frac{\tilde{\alpha}_{\mathrm{tr}}}{\hbar}\langle\{\hat{\bold{\sigma}},
\hat{\mathrm{H}}\} \rangle  + 2\frac{\alpha_{\mathrm{tr}} +
  \alpha_{\mathrm{l}} }{\hbar} \langle \hat{\mathrm{H}} \rangle
\langle \hat{\bold{\sigma}} \rangle  
  \;. 
\end{eqnarray} 
In the following we assume that the Hamiltonian $\hat{\mathrm{H}}$ is
given by: 
\begin{equation}
\hat{\mathrm{H}} = -\frac{g\mu_B}{2} \mathbf{B} \cdot
\hat{\bold{\sigma}} \;.
\end{equation}
Now, lets look for the terms on the right hand side and take into account
that $\gamma = g\mu_B/\hbar$:

The first term describes the precessional motion:
\begin{eqnarray}
-\frac{i}{\hbar}\langle[\hat{\bold{\sigma}},\hat{\mathrm{H}}]\rangle =
\frac{i \gamma}{2} \langle[\hat{\bold{\sigma}}, \mathbf{B} \cdot
\hat{\bold{\sigma}} ]\rangle \;.
\end{eqnarray}
This is an vector, therefore it is more convenient to look for the
components $u,v,w \in \{x,y,z\}$ separately: 
\begin{eqnarray}
\frac{i \gamma }{2} B^v \langle[\hat{\sigma}_u,
  \hat{\sigma}_v ]\rangle = \frac{2 i^2 }{2} \gamma B^v
\epsilon_{uvw} \langle \hat{\sigma}_w \rangle =
\gamma (\mathbf{B} \times \langle \hat{\bold{\sigma}} \rangle)_u \;.  
\end{eqnarray} 
Here, we have used the commutator relation $[\hat{\sigma}_u,\hat{\sigma}_v] =
2i\epsilon_{uvw}\hat{\sigma}_w$ and the Einstein sum convention after
which we have to sum over identical indices. We find finally, written
as vector :   
\begin{eqnarray}
-\frac{i}{\hbar}\langle[\hat{\bold{\sigma}},\hat{\mathrm{H}}]\rangle =
\gamma \mathbf{B} \times \langle \hat{\bold{\sigma}} \rangle \;.   
\end{eqnarray} 

The second term is:
\begin{eqnarray}
-\frac{\tilde{\alpha}_{\mathrm{tr}}}{\hbar}\langle\{\hat{\bold{\sigma}},
\hat{\mathrm{H}}\} \rangle = \frac{\gamma \alpha_{\mathrm{tr}}}{2}
\langle\{\hat{\bold{\sigma}}, \mathbf{B} \cdot \hat{\bold{\sigma}} \}
\rangle \langle \psi | \psi \rangle \;. 
\end{eqnarray}
Again, this is a vector and it is more convenient to look for one
component:
\begin{eqnarray}
\frac{\gamma \alpha_{\mathrm{tr}}}{2} B^v \langle\{\hat{\sigma}_u,
\hat{\sigma}_v \} \rangle \langle \psi | \psi \rangle = \gamma
  \alpha_{\mathrm{tr}} B_u [\langle \psi | \psi \rangle]^2 \;.
\end{eqnarray}
Here, we have used the anti-commutator relation $\{\hat{\sigma}_u,
\hat{\sigma}_v \} = 2 \delta_{uv} \hat{\mathbf{1}}$ and $\langle
\hat{\mathbf{1}} \rangle = \langle \psi | \hat{\mathbf{1}} |\psi
\rangle =  \langle \psi | \psi \rangle$. Written as vector:
\begin{eqnarray}
-\frac{\tilde{\alpha}_{\mathrm{tr}}}{\hbar}\langle\{\hat{\bold{\sigma}},
\hat{\mathrm{H}}\} \rangle = \gamma \alpha_{\mathrm{tr}} \mathbf{B}
    [\langle \psi | \psi \rangle]^2 \;. 
\end{eqnarray}
Introducing the vector $\tilde{\mathbf{m}}$ with
$\tilde{\mathbf{m}}^2 =  \tilde{\mathbf{m}} \cdot \tilde{\mathbf{m}} =
1$ and the relation $\tilde{\mathbf{m}} \langle \psi | \psi \rangle = \langle
\hat{\mathbf{\sigma}} \rangle$, and $\langle \hat{\bold{\sigma}}
\rangle^2 = \langle \hat{\bold{\sigma}}\rangle \cdot \langle
\hat{\bold{\sigma}}\rangle = [\langle \psi | \psi \rangle]^2 \leq 1$
we get: 
\begin{eqnarray}
-\frac{\tilde{\alpha}_{\mathrm{tr}}}{\hbar}\langle\{\hat{\bold{\sigma}},
\hat{\mathrm{H}}\} \rangle = \gamma \alpha_{\mathrm{tr}} \mathbf{B}
\tilde{\mathbf{m}}^2 [\langle \psi | \psi \rangle]^2 = \gamma
\alpha_{\mathrm{tr}} \mathbf{B} (\langle \hat{\bold{\sigma}}\rangle
\cdot \langle \hat{\bold{\sigma}}\rangle ) \;. 
\end{eqnarray}
In the case $\alpha_{\mathrm{l}} = 0$ (no longitudinal relaxation) we have
$\langle \psi | \psi \rangle = 1$ and therefore $\langle
\hat{\bold{\sigma}} \rangle^2 = 1$. However, this is not the case for
$\alpha_{\mathrm{l}} \neq 0$. Here, we have $0 \leq \langle \psi |
\psi \rangle \leq 1$. Therefore, during the calculation it was needed
to use $[\langle \psi | \psi \rangle]^2$: one $\langle \psi | \psi
\rangle$ came from the expectation value of the anti-commutator
$\langle \{\hat{\sigma}_u,\hat{\sigma}_v \}\rangle = 2 \delta_{uv}
\langle \psi | \psi \rangle$ and the second from the modification:
$\tilde{\alpha}_{\mathrm{tr}} = \alpha_{\mathrm{tr}} \langle \psi |
\psi \rangle$. This means that the modification we have introduced in
the paper is also needed here to get the correct result in the case
$\alpha_{\mathrm{l}} \neq 0$. 

The last term is given by:
\begin{eqnarray}
2\frac{\alpha_{\mathrm{tr}} + \alpha_{\mathrm{l}} }{\hbar} \langle
\hat{\mathrm{H}} \rangle \langle \hat{\bold{\sigma}} \rangle = -\gamma
(\alpha_{\mathrm{tr}} + \alpha_{\mathrm{l}}) (\mathbf{B} \cdot \langle
\hat{\bold{\sigma}} \rangle) \langle \hat{\bold{\sigma}} \rangle \;.
\end{eqnarray}  
If we combine this (only the transversal relaxation, meaning only the
part with $\alpha_{\mathrm{tr}}$) with the second term we find:
\begin{eqnarray}
\gamma \alpha_{\mathrm{tr}} \Big(\mathbf{B} (\langle \hat{\bold{\sigma}}\rangle
\cdot \langle \hat{\bold{\sigma}}\rangle ) -(\mathbf{B} \cdot \langle
\hat{\bold{\sigma}} \rangle) \langle \hat{\bold{\sigma}} \rangle
\Big) = -\gamma \alpha_{\mathrm{tr}} \langle \hat{\bold{\sigma}}
\rangle \times \Big( \langle \hat{\bold{\sigma}} \rangle \times
\mathbf{B} \Big) \;,
\end{eqnarray}
and in total:
\begin{eqnarray}
\frac{\mathrm{d}\langle\hat{\bold{\sigma}}\rangle}{\mathrm{d}t} =
\gamma \mathbf{B} \times \langle \hat{\bold{\sigma}} \rangle -\gamma
\alpha_{\mathrm{tr}} \langle \hat{\bold{\sigma}} 
\rangle \times \Big( \langle \hat{\bold{\sigma}} \rangle \times
\mathbf{B} \Big) -\gamma \alpha_{\mathrm{l}} \Big(\mathbf{B} \cdot \langle
\hat{\bold{\sigma}} \rangle \Big) \langle \hat{\bold{\sigma}} \rangle \;.
\end{eqnarray}

\section{Spin density operator for $S = 1/2$} \label{s:DensityOperator}
The wave function of a single spin $S = 1/2$ is given by:
\begin{eqnarray}
|\psi\rangle  = \cos\frac{\theta}{2} |\uparrow \rangle +
|\sin\frac{\theta}{2} e^{i\phi} | \downarrow \rangle \,.
\end{eqnarray}
The same is true for the $S = 1/2$ coherent spin state:
\begin{eqnarray}
|\psi\rangle &=&
e^{-\frac{i\phi\hat{S}_z}{\hbar}}e^{-\frac{i\theta\hat{S}_y}{\hbar}}
|\uparrow\rangle \nonumber \\ 
&=& \cos\frac{\theta}{2} |\uparrow \rangle + \sin\frac{\theta}{2} e^{i\phi}
| \downarrow \rangle \,.
\end{eqnarray}
Then, the corresponding density operator is given by:
\begin{eqnarray}
\hat{\rho} = |\psi\rangle\langle \psi | &=& \left(\cos\frac{\theta}{2}
|\uparrow \rangle + \sin\frac{\theta}{2} e^{i\phi} | \downarrow
\rangle\right)\left(\langle \uparrow | \cos\frac{\theta}{2}  + \langle
\downarrow | \sin\frac{\theta}{2} e^{-i\phi} \right) \nonumber \\
&=& \cos^2\frac{\theta}{2} |\uparrow \rangle \langle \uparrow | +
\cos\frac{\theta}{2}\sin\frac{\theta}{2} e^{-i\phi} |\uparrow \rangle
\langle \downarrow | + \cos\frac{\theta}{2}\sin\frac{\theta}{2}
e^{i\phi} |\downarrow \rangle \langle \uparrow | +
\sin^2\frac{\theta}{2} |\downarrow \rangle \langle \downarrow | \;.
\nonumber \\
\end{eqnarray}
With $\sin(2\theta) = 2\sin\theta\cos\theta$, $\sin^2\theta = 1/2 -
1/2\cos(2\theta)$ and $\cos^2\theta = 1/2 + 1/2\cos(2\theta)$, we are
able to write:
\begin{eqnarray}
\hat{\rho} = |\psi\rangle\langle \psi | &=&
\frac{1}{2}\Big((1+\cos\theta) |\uparrow \rangle \langle \uparrow | +
\sin\theta e^{-i\phi} |\uparrow \rangle \langle \downarrow | +
\sin\theta e^{i\phi} |\downarrow \rangle \langle \uparrow | +
(1-\cos\theta) |\downarrow \rangle \langle \downarrow |
\Big)\;. \nonumber \\
\end{eqnarray}
With $|\uparrow \rangle = \left(\begin{array}{c} 1 \\ 0 \end{array} \right)$,
and $|\downarrow \rangle = \left(\begin{array}{c} 0 \\ 1 \end{array} \right)$
we can write this as matrix:
\begin{eqnarray}
\hat{\rho} = \frac{1}{2} \left(\begin{array}{cc} 1+\cos\theta &
  \sin\theta e^{-i\phi} \\ \sin\theta e^{i\phi} &
  1-\cos\theta \end{array}\right) \;. 
\end{eqnarray}
With $\exp(\pm i\phi) = \cos\phi \pm i \sin\phi$, $\langle
\hat{\sigma}_x \rangle = \sin\theta \cos\phi$, $\langle \hat{\sigma}_y
\rangle = \sin\theta \sin\phi$, and $\langle \hat{\sigma}_z \rangle =
\cos\theta$ we get finally:
\begin{eqnarray}
\hat{\rho} = \frac{1}{2} \left(\begin{array}{cc} 1+ \langle
\hat{\sigma}_z \rangle & \langle \hat{\sigma}_x \rangle - i \langle
\hat{\sigma}_y \rangle \\ \langle \hat{\sigma}_x \rangle + i \langle
\hat{\sigma}_y \rangle  & 1-\langle \hat{\sigma}_z
\rangle \end{array}\right) \;. 
\end{eqnarray}
This can be written in a more compact form as:
\begin{eqnarray}
\hat{\rho} = \frac{1}{2} \left(\hat{\mathbf{1}} + \langle
\hat{\bold{\sigma}} \rangle \cdot \hat{\bold{\sigma}} \right) \;,
\end{eqnarray}
with the Pauli matrices $\hat{\sigma}_\eta$, $\eta \in \{x,y,z\}$, and
$\hat{\mathbf{1}}$ the identity matrix. 

\section{Derivation of the qLLB Eq. using density operator for an
  arbitrary spin quantum number $S$} \label{s:qLLBDensity}       
 
The density operator for any spin quantum number $S$ can be written
as \cite{pokrovskyPRB03}:
\begin{eqnarray} \label{density}
\hat{\rho} =
  \frac{1}{2S+1} \hat{\mathbf{1}} + \mathbf{m}
  \cdot \hat{\mathbf{S}} + \mathrm{higher}\;\mathrm{order}\;\mathrm{tensors}\;,
\end{eqnarray}
with 
\begin{eqnarray} \label{density2}
\mathbf{m} = \frac{\langle \hat{\mathbf{S}} \rangle}{\hbar S} =
\mathbf{e}_S \;,
\end{eqnarray}
where $\mathbf{e}_S$ is a normalized vector:
\begin{eqnarray}
\mathbf{e}_S = \left(\begin{array}{c} \cos\phi \sin\theta \\ \sin\phi
  \sin\theta \\ \cos\theta \end{array} \right) \;.
\end{eqnarray}
The number and appearance of the higher order tensor terms is given by the
spin quantum number $S$: Such terms appear only for $S > 1/2$
\cite{hofmannPRA04,kimuraPLA03}. Furthermore, the number of higher order tensor
terms increases with increasing $S$. For the conventional spin
dynamics the vector terms [second term in Eq.~(\ref{density})] are
important. The higher order tensor terms will lead to separate
differential equations which don't interfere with the dynamics of the
Bloch vector $\mathbf{m}$. 

Now, the quantum Landau-Lifshitz-Bloch (qLLB) equation is defined as:
\begin{equation} \label{TDSELLB}
i\hbar\frac{\mathrm{d}}{\mathrm{d}t} | \psi\rangle =
\left(\hat{\mathrm{H}} - 
  i\alpha_{\mathrm{tr}}[\langle \psi | \psi \rangle
  \hat{\mathrm{H}} - 
  \langle\hat{\mathrm{H}}\rangle]  + 
  i\alpha_{\mathrm{l}}  
    \langle\hat{\mathrm{H}}\rangle \right)  |\psi \rangle\;,
\end{equation}
and the corresponding von Neumann equation is:
\begin{equation} \label{Liouville}
\frac{\mathrm{d}\hat{\rho}}{\mathrm{d}t} =
\frac{i}{\hbar}[\hat{\rho},\hat{\mathrm{H}}] -
  \frac{\alpha_{\mathrm{tr}}}{\hbar}[\hat{\rho},[\hat{\rho},\hat{\mathrm{H}}]]
  + \frac{\alpha_{\mathrm{l}}}{\hbar}
  (\hat{\rho}\hat{\mathrm{H}})\hat{\rho}   \;.
\end{equation} 
In both differential equations (qLLB and von Neumann equation) the
terms on the right hand side correspond to a precessional motion
(first term), transversal (second term) and longitudinal relaxation
(last term). In the von Neumann equation within the publication the
longitudinal relaxation term is defined with an additional factor two
which has been introduced to make a symmetric decoupling going from
the von Neumann equation to the TDSE. However, this term only scales
the strength of the relaxation and can be skipped because the function
$\alpha_{\mathrm{l}}$ has to be defined anyhow. Furthermore, the
decoupling of the von Neumann equation to derive the TDSE can be also
asymmetric meaning that only the one differential equation contains
the longitudinal relaxation term which becomes after skipping $\langle
\psi |$ the searched TDSE while the corresponding conjugate complex TDSE
does not contain this term.  

In the following we have assumed that the Hamilton operator
$\hat{\mathrm{H}}$ is given by: 
\begin{eqnarray} \label{Ham}
\hat{\mathrm{H}} = -\gamma \mathbf{B} \cdot \hat{\mathbf{S}}\;,
\end{eqnarray}
with $\mathbf{B}$ an effective field.

Inserting this equation into Eq.~(\ref{Liouville}) together with
Eq.~(\ref{density}) and (\ref{Ham}) we find the following differential
equation:
\begin{equation} \label{QMLLB}
\frac{\partial 
  \mathbf{m}}{\partial t} = \gamma \mathbf{m} \times
\mathbf{B} - \gamma \tilde{\alpha}_{\mathrm{tr}}
\mathbf{m} \times  \left( \mathbf{m} \times \mathbf{B}
\right) - \gamma \tilde{\alpha}_{\mathrm{l}} \left( \mathbf{m} \cdot
\mathbf{B}\right) \mathbf{m} \;.
\end{equation}
In detail: For the left hand side we immediately find:
\begin{equation}
\frac{\partial \hat{\rho}}{\partial t} = \frac{\partial
  \mathbf{m}}{\partial t} \cdot  \hat{\mathbf{S}} \;.
\end{equation}
  
Now, from the connection between classical physics and quantum
mechanics we know that the commutator $[\hat{A},\hat{B}]$ and the
Poisson bracket $\{A,B\}$ are connected by
$[\hat{A},\hat{B}]\;\leftrightarrow\;i\hbar\{A,B\}$. 
Then, the Poisson bracket of a classical spin system is given by
\cite{lakshmananPTRS11,wieserEPJB15}: 
\begin{eqnarray}
\{A,B\} &=& \frac{\partial A}{\partial S_n}\frac{\partial
  B}{\partial S_m} S_l \epsilon_{nml} \;,
\end{eqnarray}
where the Einstein sum convention has been used. $A$, $B$ and $C$
are functions of a spin tensor $\mathbf{S}$ of rank $n$ and   
$S_n$, $S_m$, $\ldots$, $S_v$ are the components of this
spin tensor. As mentioned before we are just interested in the spin
operator $\hat{\mathbf{S}}$ and therefore in spin tensors
$\hat{S}_n$ of first rank. This means: $S_n$, $S_m$,
$\ldots$, $S_v$ are the spin components $S_x$, $S_y$, and
$S_z$. Furthermore, $\epsilon_{nml}$ is the Levi-Civita tensor. With this
informations it is easy to calculate the
commutator $[\hat{\rho},\hat{\mathrm{H}}]$:  
\begin{eqnarray} \label{P1}
\frac{i}{\hbar}[\hat{\rho},\hat{\mathrm{H}}] &=&  - \frac{i\gamma}{\hbar}
[\mathbf{m} \cdot \hat{\mathbf{S}}, \mathbf{B} \cdot
\hat{\mathbf{S}}] = -i \hbar \frac{i}{\hbar} \gamma
\{\mathbf{m} \cdot \hat{\mathbf{S}}, \mathbf{B} \cdot
\hat{\mathbf{S}}\} = \gamma
\{\mathbf{m} \cdot \hat{\mathbf{S}}, \mathbf{B} \cdot
\hat{\mathbf{S}}\} \nonumber \\
&=& \gamma m_u B_v \frac{\partial \hat{S}_u}{\partial
  \hat{S}_n}\frac{\partial \hat{S}_v}{\partial \hat{S}_m}
\hat{S}_l \epsilon_{nml} = \gamma m_u
B_v \hat{S}_l \delta_{un}\delta_{vm} \epsilon_{nml} = \gamma
(\mathbf{m} \times \mathbf{B}) \cdot \hat{\mathbf{S}} \;. 
\end{eqnarray}
This result can be easily proved with the commutator:
\begin{equation}
[\hat{S}_u,\hat{S}_v] = i\hbar \epsilon_{uvw}\hat{S}_w \;.
\end{equation}
The commutator $[\hat{\rho},\hat{\mathrm{H}}]$ can be written as:
\begin{eqnarray} \label{P1a}
\frac{i}{\hbar}[\hat{\rho},\hat{\mathrm{H}}] &=&  - \frac{i \gamma}{\hbar}
[\mathbf{m} \cdot \hat{\mathbf{S}}, \mathbf{B} \cdot
\hat{\mathbf{S}}] = - \frac{i \gamma}{\hbar} m_u B_v
[\hat{S}_u,\hat{S}_v] = \gamma m_u B_v \hat{S}_w \epsilon_{uvw}
 = \gamma \left(\mathbf{m} \times \mathbf{B}
  \right) \cdot \hat{\mathbf{S}} \;, \nonumber \\
\end{eqnarray}
where with the definition of the vector product via Levi-Civita tensor it is
easy to reproduce the above result without using the Poisson bracket.

The next term is the transversal relaxation term. Here, we have the
double commutator:
\begin{eqnarray} \label{P2}
\frac{\alpha_{\mathrm{tr}}}{\hbar}[\hat{\rho},[\hat{\rho},\hat{\mathrm{H}}]] =
-i\alpha_{\mathrm{tr}}\gamma [\mathbf{m} \cdot
  \hat{\mathbf{S}}, \mathbf{v} \cdot \hat{\mathbf{S}}] \;.
\end{eqnarray}
Here, we have used the fact that we have already solved the
commutator:
\begin{eqnarray} \label{P1b}
[\hat{\rho},\hat{\mathrm{H}}] = -i \hbar \gamma \left(\mathbf{m}
\times \mathbf{B} \right) \cdot \hat{\mathbf{S}} = -i \hbar
\gamma \mathbf{v} \cdot \hat{\mathbf{S}}\;. 
\end{eqnarray}
Then the corresponding commutator with spin operators
$\hat{\mathbf{S}}$ is given by :
\begin{eqnarray}
-i\alpha_{\mathrm{tr}} \gamma [\mathbf{m} \cdot \hat{\mathbf{S}}, 
\mathbf{v} \cdot \hat{\mathbf{S}}  ] =  \alpha_{\mathrm{tr}} \gamma
\hbar \left(\mathbf{m} \times \mathbf{v}\right) \cdot \hat{\mathbf{S}} =
\alpha_{\mathrm{tr}} \gamma \hbar \left(\mathbf{m} \times \left(\mathbf{m}
\times \mathbf{B} \right) \right)\cdot \hat{\mathbf{S}}   \;,
\end{eqnarray}
and therefore:
\begin{eqnarray} \label{P2Fin}
\frac{\alpha_{\mathrm{tr}}}{\hbar}[\hat{\rho},[\hat{\rho},\hat{\mathrm{H}}]] &=&
\alpha_{\mathrm{tr}} \hbar \left(\mathbf{m} \times \left(\mathbf{m} \times
\mathbf{B} \right)
\right) \cdot \hat{\mathbf{S}} \;.
\end{eqnarray}

Finally, the longitudinal relaxation term is given by:
\begin{eqnarray} 
\frac{\alpha_{\mathrm{l}}}{\hbar}(\hat{\rho}\hat{\mathrm{H}})\hat{\rho}
= -\frac{\alpha_{\mathrm{l}} \gamma}{\hbar} \left((\mathbf{m} 
\cdot \hat{\mathbf{S}})(\mathbf{B} \cdot 
\hat{\mathbf{S}})\right) \mathbf{m} \cdot \hat{\mathbf{S}} \;.
\end{eqnarray}
Here, we have already skipped the noncontributing term with the
identity operator $\hat{\mathbf{1}}$. With $\cal{M} = \mathbf{m}
\cdot \hat{\mathbf{S}}$, $\cal{B} = \mathbf{B}
\cdot \hat{\mathbf{S}}$, and the definition of the inner product using
the trace we have:
\begin{eqnarray} 
\frac{\alpha_{\mathrm{l}}}{\hbar}(\hat{\rho}\hat{\mathrm{H}})\hat{\rho}
= - \frac{\alpha_{\mathrm{l}} \gamma}{\hbar} \langle {\cal M},{\cal B}
\rangle \mathbf{m} \cdot \hat{\mathbf{S}} = 
- - \frac{\alpha_{\mathrm{l}} \gamma}{N \hbar} \mathrm{Tr}\left({\cal
  M}{\cal B}\right) \mathbf{m} \cdot \hat{\mathbf{S}}  \;,  
\end{eqnarray}
where $N = 2S+1$ is the normalization because the inner product does
not necessarily deliver normalized results. In the case $S = 1/2$ the
normalization was not needed due to the fact that there the additional
$1/2$ factors corresponding the transformation $\hat{\mathbf{S}} =
(\hbar/2) \hat{\mathbf{\sigma}}$ have done the job and the inner product
was normalized. In general this is not the case and therefore a
normalization needed. It seems that the additional factor two in front
of the longitudinal relaxation term works fine with $S = 1/2$ but not
with $S > 1/2$.

Now, with aid of the geometric product (Grassmann):
\begin{eqnarray}
{\cal M}{\cal B} &=& {\cal M}\cdot{\cal B} + {\cal M}\wedge {\cal B}
\nonumber \\
&=& \left(\mathbf{m}\cdot\mathbf{B}\right) \hat{\mathbf{1}} + i
\left(\mathbf{m}\times\mathbf{B}\right) \cdot \hat{\mathbf{S}} \;,
\end{eqnarray}
it is easy to calculate the trace using:
\begin{eqnarray}
\mathrm{Tr}\left(\hat{\mathbf{1}}\right) = 2S+1 \;\;\; \mathrm{and}
\;\;\; \mathrm{Tr}\left(\hat{\mathbf{S}}\right) = 0 \;.
\end{eqnarray} 
Therefore the trace is given by:
\begin{equation}
\frac{1}{2S+1}\mathrm{Tr}\left({\cal M}{\cal B}\right) =
\frac{1}{2S+1}\left[ \left(\mathbf{m}\cdot\mathbf{B}\right)
\mathrm{Tr}\left(\hat{\mathbf{1}}\right) + i 
\left(\mathbf{m}\times\mathbf{B}\right)_n
\mathrm{Tr}\left(\hat{S}_n\right)\right] =  
\mathbf{m}\cdot\mathbf{B} \;.
\end{equation}
Finally, we find:
\begin{eqnarray} \label{P3}
\frac{\alpha_{\mathrm{l}}}{\hbar}(\hat{\rho}\hat{\mathrm{H}})\hat{\rho} =
- \frac{\alpha_{\mathrm{l}} \gamma}{\hbar}
\left(\mathbf{m}\cdot\mathbf{B}\right) \mathbf{m} \cdot 
\hat{\mathbf{S}} \;.
 \end{eqnarray}

Bringing all parts together and skipping the
$\hat{\mathbf{S}}$ on the right hand side in any term and setting
$\alpha_{\mathrm{tr}}\hbar = \tilde{\alpha}_{\mathrm{tr}}$ as well
$\alpha_{\mathrm{l}}/\hbar = \tilde{\alpha}_{\mathrm{l}}$ we get
Eq.~(\ref{QMLLB}) as final result. Please notice that both
$\tilde{\alpha}_{\mathrm{tr}}$ as well as
$\tilde{\alpha}_{\mathrm{l}}$ have the dimension of $\hbar$. This is
similar to the Landau-Lifshitz constant $\lambda$ which has the same 
dimension as the gyromagnetic ratio $\gamma$.    

At the end a view words more about the inner product. As mentioned
before in Euclidian vector space we can write any vector $\mathbf{a}$
as a $n\times 1$ or $1\times n$ matrix e.g. in ${\rm I\!R}^3$ as
$\mathbf{a} = (a_x,a_y,a_z)^T$. Alternative we can develop the vector
with respect to its basis vectors $\hat{\mathbf{x}} = (1,0,0)^T$,
$\hat{\mathbf{y}} = (0,1,0)^T$, and $\hat{\mathbf{z}} =(0,0,1)^T$:
$\mathbf{a} = a_z\hat{\mathbf{z}} + a_z\hat{\mathbf{z}} +
a_z\hat{\mathbf{z}}$. In the case of the von Neumann equation we are
dealing with matrices ${\cal M} = \mathbf{m}\cdot\hat{\mathbf{S}}$,
and ${\cal B} = \mathbf{B}\cdot\hat{\mathbf{S}}$ which can be also
understood as vectors. Here the spin operators $\hat{S}_x$,
$\hat{S}_y$, $\hat{S}_z)^T$ have the same function as the basis
vectors $\hat{\mathbf{x}}$, $\hat{\mathbf{y}}$, and
$\hat{\mathbf{z}}$: ${\cal M} = m_x \hat{S}_x + m_y \hat{S}_y + m_z
\hat{S}_z$, and ${\cal B} = B_x \hat{S}_x + B_y \hat{S}_y + B_z
\hat{S}_z$. Similar to the Euclidian vector we can also write this as
a $n\times 1$ or $1\times n$ matrix: e.g. ${\cal M} = (m_x,m_y,m_z)^T$, and
${\cal B} = (B_x,B_y,B_z)^T$. With this we can write:
\begin{eqnarray}
{\cal M}{\cal B} =
(\mathbf{m}\cdot\hat{\mathbf{S}})(\mathbf{B}\cdot\hat{\mathbf{S}}) = 
  (m_x,m_y,m_z) \left(\begin{array}{c}B_x\\B_y\\B_z \end{array}
\right) = \mathbf{m}\cdot \mathbf{B} \;,
\end{eqnarray}  
and therefore:
\begin{eqnarray} 
\frac{\alpha_{\mathrm{l}}}{\hbar}(\hat{\rho}\hat{\mathrm{H}})\hat{\rho}
= -\frac{\alpha_{\mathrm{l}} \gamma}{\hbar} \left({\cal M}{\cal
  B}\right) {\cal M} = -\frac{\alpha_{\mathrm{l}} \gamma}{\hbar}
\left(\mathbf{m}\cdot \mathbf{B} \right) \mathbf{m} \cdot
\hat{\mathbf{S}} \;.
\end{eqnarray}
With this, it also becomes more clear why we need a normalization of
the inner product using the trace: While the Euclidian basis
vectors $\hat{\mathbf{x}}$, $\hat{\mathbf{y}}$, and
$\hat{\mathbf{z}}$ are normalized it is not the case for the spin
operator $\hat{S}_x$, $\hat{S}_y$, and $\hat{S}_z$. Moreover the size
of this matrices change with changing spin quantum number $S$. The
same is true for the corresponding identity matrix $\hat{\mathbf{1}}$.
Therefore the norm $N = 2S+1$ has to be adapted to $S$.


\bibliography{Cite}